\documentclass[aps,prl,floatfix,nofootinbib,superscriptaddress,twocolumn,footinbib]{revtex4-1}
\usepackage{graphicx}  
\usepackage{dcolumn}   
\usepackage{bm}        
\usepackage{amssymb}   
\usepackage{amsmath}
\usepackage{comment}
\usepackage{graphicx}
\usepackage{color}
\usepackage{slashed}
\usepackage{hyperref}
\usepackage[dvipsnames]{xcolor}

\usepackage{xr}
\newcommand{\Eq}[1]{Eq.(\ref{#1})}

\newcommand{\Fig}[1]{Fig. (\ref{#1})}

\newcommand{\tsubsat}{t_{\rm sat}}
\newcommand{\s}{{\mathbf{q},h}}
\newcommand{\sprime}{{\mathbf{q}',h'}}

\begin{document}

\author{Mark Mace}
\email{mark.f.mace@gmail.com} 
\affiliation{Department of Physics, University of Jyv\"{a}skyl\"{a}, P.O. Box 35, 40014, Jyv\"{a}skyl\"{a}, Finland}
\affiliation{Helsinki Institute of Physics, University of Helsinki, P.O. Box 64, 00014, Helsinki, Finland}
\author{Niklas Mueller}
\email{nmueller@bnl.gov}
\affiliation{Physics Department, Brookhaven National Laboratory, Bldg. 510A, Upton, NY 11973, USA}
\author{S\"{o}ren Schlichting}
\email{sschlichting@physik.uni-bielefeld.de}
\affiliation{Fakult\"{a}t f\"{u}r Physik, Universit\"{a}t Bielefeld, D-33615 Bielefeld, Germany}
\author{Sayantan Sharma}
\email{sayantans@imsc.res.in}
\affiliation{The Institute of Mathematical Sciences, HBNI, Chennai 600113, India}

\title{Chiral instabilities \& the onset of chiral turbulence in QED plasmas}
\date{\today}
\begin{abstract} 
We present a first principles study of chiral plasma instabilities and the onset of chiral turbulence in QED plasmas with strong gauge matter interaction ($e^2N_{ f}=64$), far from equilibrium. By performing classical-statistical lattice simulations of the microscopic theory, we show that the generation of strong helical magnetic fields from a helicity imbalance in the fermion sector proceeds via three distinct phases. During the initial linear instability regime the helicity imbalance of the fermion sector causes an exponential growth(damping) of magnetic field modes with right (left) handed polarization, for which we extract the characteristic growth (damping) rates. 
Secondary growth of unstable modes accelerates the helicity transfer from fermions to gauge fields and ultimately leads to the emergence of a self-similar scaling regime characteristic of a decaying turbulence, where magnetic helicity is efficiently transferred to macroscopic length scales. Within this turbulent regime, the evolution of magnetic helicity spectrum can be described by an infrared power-spectrum with spectral exponent $\kappa = 10.2\pm 0.5$ and dynamical scaling exponents $\alpha=1.14\pm 0.50$ and $\beta=0.37\pm 0.13$.
 \end{abstract}
\maketitle

Novel macroscopic phenomena related to the in- and out-of-equilibrium dynamics of chiral fermions have inspired a significant amount of theoretical and experimental developments in recent years. By means of the Chiral Magnetic Effect (CME) ~\cite{Kharzeev:2007jp,Fukushima:2008xe,Kharzeev:2012ph,Kharzeev:2015znc,Skokov:2016yrj,Avkhadiev:2017fxj,Hirono:2018bwo,Becattini:2013vja}, one hopes  for example, to investigate the topological structure of Quantum Chromodynamics (QCD) in ultra-relativistic heavy ion collisions \cite{Klinkhamer:1984di,Dashen:1974ck}; or explore new kinds of transport phenomena in condensed matter systems \cite{Hosur:2013kxa,Li:2014bha,Mizher:2016mfq,Gorbar:2017lnp}, including dissipationless electric transport in Dirac- and Weyl-semimetals \cite{Son:2012bg,Li:2014bha,Huang:2015eia}, as well as applications to opto-electronics \cite{yudin2015dynamics,Chan:2015dwa,Kaushik:2018tjj}.

One important aspect of anomalous transport concerns the question how chirality is transferred between gauge fields and fermionic degrees of freedom. While in QCD plasmas, chirality transfer can be efficiently accomplished by sphaleron transitions between different topological sectors of the non-Abelian gauge theory~\cite{McLerran:1990de},  the situation is markedly different in Abelian plasmas. Quantum Electrodynamics (QED) is topologically trivial and a different mechanism has to be invoked to convert fermionic chirality into magnetic helicity (and vice versa). In this context,  a novel type of `chiral' plasma instability has been suggested as a viable mechanism, whereby a chirality imbalance in the fermion sector can generate helical magnetic fields that exist on macroscopic length scales \cite{Vilenkin:1980fu,Akamatsu:2013pjd}.

Such effects have been proposed as a possible explanation for the creation of large scale helical magnetic fields in astrophysical systems, such as supernovae and compact stars~\cite{Ohnishi:2014uea,Grabowska:2014efa,Kaminski:2014jda,Kaplan:2016drz,Sen:2016jzl,Anber:2016yqr}, or in primordial plasmas of the early universe~\cite{Son:1998my,Boyarsky:2011uy,Boyarsky:2012ex,Brandenburg:2017rcb,Kharzeev:2019rsy} where the interplay between fermion chirality and magnetic helicity could be responsible for the transport of magnetic helicity from microscopic to macroscopic scales.

Chiral instabilities in QED plasmas have been studied previously based on different theoretical approaches including magneto-hydrodynamics (MHD) \cite{Campanelli:2007tc,Brandenburg:2014mwa,Tuchin:2014iua,Pavlovic:2016gac,Yamamoto:2016xtu,Rogachevskii:2017uyc,Schober:2017cdw,DelZanna:2018dyb}, compact and non-compact lattice QED simulations~\cite{Buividovich:2015jfa,Figueroa:2017hun,Figueroa:2017qmv,Figueroa:2019jsi}, kinetic theory {\cite{Manuel:2015zpa,Dvornikov:2015ena}}, linearized perturbation theory \cite{Kojima:2019eed} and effective action approaches~\cite{Grabowska:2014efa,Kaplan:2016drz}. Despite strong theoretical interest, previous studies have come to different conclusions, and there appears to be no general agreement regarding the detailed mechanisms and viability of such a scenario. 

In this article we present a comprehensive study of chiral instabilities using microscopic real-time lattice simulations of strongly-coupled QED plasmas. Starting from a helicity imbalance in the fermion sector, we employ a classical-statistical description \cite{Polkovnikov:2009ys,Berges:2012ev,Kurkela:2012hp,Jeon:2013zga,Berges:2015kfa} to simulate the subsequent non-equilibrium evolution of the system from first principles. We demonstrate that chiral instabilities in QED-like theories follow a characteristic pattern of quantum many body systems subject to instabilities~\cite{Nazarenko,Micha:2004bv,Boyanovsky:1996sq,Micha:2002ey,Berges:2002cz,Berges:2013lsa}, where the exponential growth of unstable modes leads to the emergence of turbulent behavior. 
Based on our microscopic simulations, we are able to characterize the entire sequence of events, starting from the extraction of growth- (and decay-) rates of helical gauge field modes in the primary and secondary instability regimes all the way to the turbulent scaling regime, for which we extract,  for the first time, the relevant far-from-equilibrium scaling exponents.

\textit{Simulation technique.} We perform real-time simulations of $N_f$ degenerate flavors of quantum Dirac fermions of mass $m$ and charge $e$, coupled to classical-statistical  $U(1)$ gauge fields~\cite{Aarts:1998td,Berges:2010zv,Kasper:2014uaa,Mueller:2016aao,Mueller:2016ven,Mace:2016shq}. By numerically solving the coupled set of Dirac equations for fermion fields $\hat{ \Psi}_\mathbf{x}(t)$
\begin{align}
\label{eq:Dirac}
i\partial_t \hat{\Psi}_{\mathbf{x}}(t)&=\gamma^{0}(-i \gamma^i D_i[A]+m)\hat{\Psi}_{\mathbf{x}}(t)\,,
\end{align}
where $D^i[A]\equiv \partial^i - ie A^i_\mathbf{x}(t)$ is the covariant derivative in temporal-axial ($A_0=0$) gauge, and Maxwell's equations for electromagnetic fields $\mathbf{E}^i_\mathbf{x}(t)$ and $\mathbf{B}^i_\mathbf{x}(t)$,
\begin{align}
\label{eq:Maxwell}
\partial_t e\mathbf{E}^i_\mathbf{x}(t) &- [\nabla \times e\mathbf{B}_\mathbf{x}(t)]^i = -e^2 N_{f} \, \mathbf{j}_{\mathbf{x}}^{i}(t)\,,
\end{align}
we include the effects of electromagnetic fields on the fermion sector in Eq.~(\ref{eq:Dirac}), as well as the non-linear back-coupling of fermion currents
$\mathbf{j}_{\mathbf{x}}^{i}(t)=\langle \frac{1}{2} [ \hat{\bar{\Psi}}_\mathbf{x}(t) \gamma^i ,\hat{ \Psi}_\mathbf{x}(t)]\rangle$, where
$\hat{\bar{\Psi}}_\mathbf{x}=\hat{{\Psi}}^\dagger_\mathbf{x}\gamma^0$,
on the dynamical evolution of the electro-magnetic fields in Eq.~(\ref{eq:Maxwell}). We note that that the classical-statistical description in Eqns.~(\ref{eq:Dirac},\ref{eq:Maxwell}) is accurate to leading order in the gauge coupling $e^2$, but to all orders in the coupling between gauge and matter fields $e^2N_f$ ~\cite{Aarts:1998td,Kasper:2014uaa}. We will specifically work in the regime of strong interactions between gauge and matter fields with $e^2N_f=64$, which is significantly larger than in single flavor QED ($e^2N_f \approx 0.09$), but necessary to resolve all relevant scales on the available size lattices.

We initially prepare the system as a chirally imbalanced charge neutral Fermi gas, by specifying the initial occupation numbers of left- (L) and right- (R) handed fermions according to a Fermi-Dirac distribution $n_F^{L/R}(t=0,\mathbf{p})=\frac{1}{e^{(E_{\mathbf{p}} \pm \mu_{h})/T}+1}$ with helicity chemical potential $\mu_h$ and energy $E_{\mathbf{p}}=\pm \sqrt{\mathbf{p}^2+m^2}$ for particles and anti-particles respectively. We focus on the low-temperature behavior $T/ \mu_h = 1/8$ and consider vacuum initial conditions for the electro-magnetic field sector, which are represented by a classical-statistical ensemble of fluctuating fields \cite{SM}. 

We discretize the theory on a $N_s^3$  spatial lattice with lattice spacing $a_s$, using a compact Hamiltonian lattice formulation of QED~\cite{Kogut:1974ag}, with $O(a_s^3)$ tree-level improved Wilson-fermions,  which as detailed in \cite{Mace:2016shq} is crucial for studying effects related to the chiral anomaly. Based on the lattice discretization, the fermion field operator $\hat{\Psi}_{\mathbf{x}}(t)$ becomes finite dimensional, and the solution to the operator equation (\ref{eq:Dirac}) can be constructed from linear combinations of a complete set of $4 N_s^3$ wave-functions \cite{Aarts:1998td,Mace:2016shq}. We employ a leap-frog solver with time step $a_t=0.001 a_s$  to solve the discretized equations of motions (\ref{eq:Dirac},\ref{eq:Maxwell}) and study different size lattices $N_s^3=32^3,48^3$ with spacings $\mu_h  a_s=2/3,1,1.25,1.5$ to monitor residual discretization effects. Simulations are performed close to the chiral limit $m \ll \mu_h$ by employing $m a_s=5\cdot 10^{-4}$ and if not stated otherwise we show results for $N_s^3=48^3$ and $\mu_h a_s=1$ expressed in terms of dimensionless quantities in units of the initial helicity chemical potential $\mu_{h}$. We will use continuum notation throughout the main text and refer to the supplemental material for a detailed description of the corresponding lattice implementation.

 \begin{figure}[tb]
\begin{center}
\includegraphics[width=0.47\textwidth]{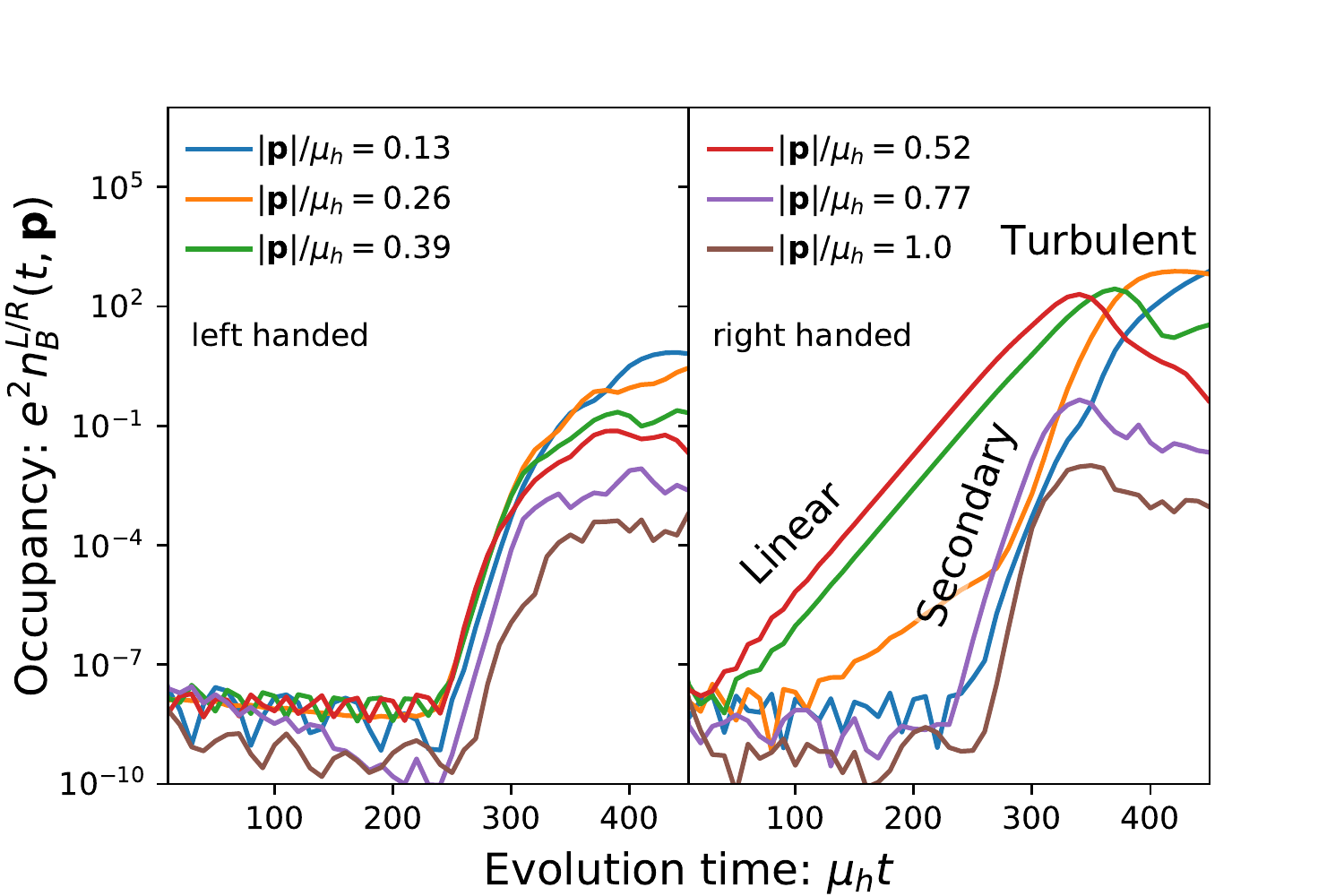}
\end{center}
\caption{Evolution of the occupation numbers  $n^{L/R}_{B}(t,\mathbf{p})$ of left handed (left panel) and right handed (right panel) magnetic field modes as a function of time $\mu_h t$ for different momenta in the range $|\mathbf{p}|/\mu_h \approx 0.1-1.0$.  Distinct regimes of linear growth, secondary growth and onset of turbulent behavior are also indicated.}
\label{fig:modes}
\end{figure}

\textit{Chiral instabilities.} Starting from an initial helicity imbalance $(\mu_h>0)$ in the fermion sector, the chiral plasma instability triggers an exponential growth of gauge fields with right-handed (circular) polarization. Separating the magnetic field into left and right handed components according to their helicity projection in Fourier space~\footnote{We use $B(t,\mathbf{p})=\frac{1}{\sqrt{V}} \int d^3\mathbf{x} ~ B_{\mathbf{x}}(t) e^{-i \mathbf{p}\cdot\mathbf{x}}$.}
\begin{align}
\mathbf{B}^{L/R}(t,\mathbf{p})= \frac{|\mathbf{p}|  \pm i\mathbf{p} \times }{2|\mathbf{p}|} \mathbf{B}(t,\mathbf{p})\,,
\end{align} 
one finds that at early times only the right-handed components experience exponential growth within a narrow momentum range; all other modes show a damped oscillatory behavior as can be qualitatively expected from the interplay of electric-magnetic fields and currents in a conducting medium. 
This is shown in  Fig.~\ref{fig:modes}, where we present the evolution of the occupation numbers of left and right handed components of the magnetic fields 
\begin{align}
n^{L/R}_{B}(t,\mathbf{p})\equiv {| \mathbf{B}^{L/R}(t,\mathbf{p}) |^2}/{|\mathbf{p}|}\,,
\end{align}
for selected momentum modes with $|\mathbf{p}|/\mu_h = 0.1-1.0$.
%

\begin{figure}[tb]
\begin{center}
\includegraphics[width=0.45\textwidth]{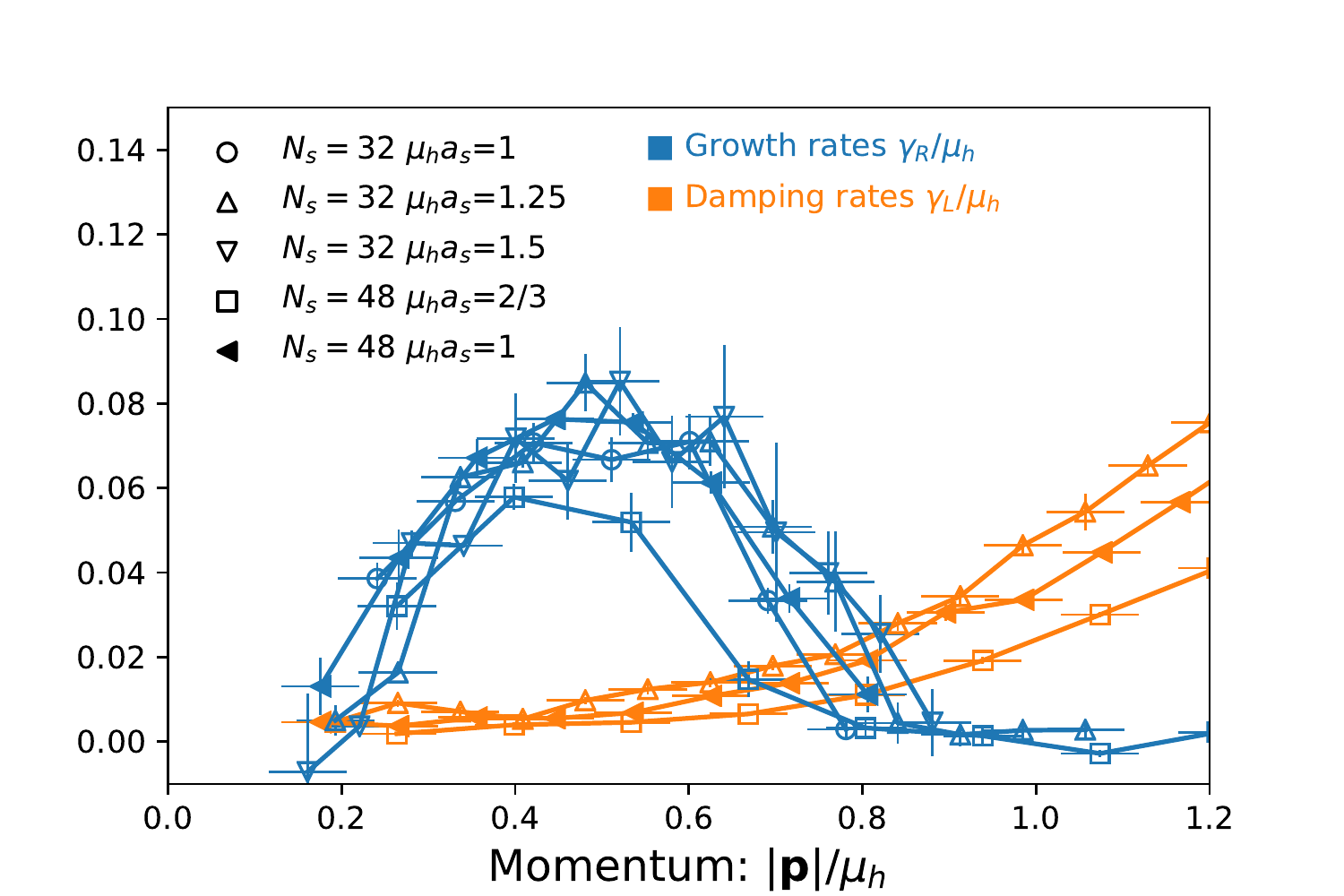}
\end{center}
\caption{Growth rates  of right-handed (blue symbols) and damping rates of left-handed (orange symbols) magnetic field modes in the linear instability regime.}
\label{fig:growthrates}
\end{figure}

We further quantify this behavior in Fig.~\ref{fig:growthrates}, where we show the growth rates (blue symbols) for the primary unstable and damping rates (orange symbols) of initially stable modes, extracted from an exponential fit to the evolution of the occupation number $n^{L/R}_{B}(t,\mathbf{p})\propto \exp{(\mp\gamma_{L/R}(|\mathbf{p}|) t)}$ in the linear instability regime. Details of the fitting procedure are given in the supplemental material~\cite{SM}.
Shown results for the growth and damping rates are quantitatively consistent across different lattices, with a maximal growth rate $\gamma_{0} \approx 0.07 \mu_{h}$ for right handed modes with momenta $|\mathbf{p}| \approx 0.5 \mu_h$. 

While the exponential growth (damping) of right-(left-)handed modes sets in almost directly after a short delay of $\mu_{h} t \approx 20-50$ due to the initial quench \footnote{Similar findings have been reported in the context of Hard-loop simulations of (Chromo-) Weibel instabilities in anisotropic plasmas, where the delay can be attributed to the formation of current fluctuations and removed by a proper choice of the initial conditions~\cite{Rebhan:2009ku,Attems:2012js}.}, the evolution continues in this fashion until non-linear interactions between unstable modes induce secondary instabilities~\cite{Berges:2002cz,Berges:2012cj}. During this second phase, which in Fig.~\ref{fig:modes} occurs around $\mu_h t\approx 250$, a large range of left and right handed momentum modes starts to exhibit exponential growth with strongly enhanced growth rates $\gamma_{\rm secondary} \sim (2-3) \gamma_{0}$, until around $ \mu_h \tsubsat  \approx 300$ the instability saturates and the exponential growth terminates. 

\textit{Energy \& Helicity transfer.}
Before we describe the dynamics at later times in more detail, it is insightful to investigate how the conserved quantities are shared and transferred between fermions and gauge field throughout the evolution of the system. Clearly, the total energy density is conserved and can be separated into the contributions from the electro-magnetic fields 
$\varepsilon_{g}(t)=\int_{V}[\frac{\mathbf{E}^2_\mathbf{x}(t)}{2} + \frac{\mathbf{B}^2_\mathbf{x}(t)}{2}]$
and the fermion sector 
$\varepsilon_{f}(t)=N_{f} \int_{V}\langle \frac{1}{2} [\hat{\bar{\Psi}}_\mathbf{x}(t),\gamma^{0} (-i \gamma^i D_i[A]+m) \hat{\Psi}_\mathbf{x}(t)  ]\rangle$,
where $\int_{V}=\frac{1}{V} \int d^3\mathbf{x}$ denotes volume averages.
By means of the axial anomaly relation \cite{Adler:1969gk,Bell:1969ts}
\begin{align}
\label{eq:anomalybudget}
\partial_t n_5(t) &= -2N_f \partial_{t} n_{h}(t) \\
& +2 im \, N_f \int_{V} \left\langle \frac{1}{2} \left[ \hat{\bar{\Psi}}_\mathbf{x}(t), \gamma^{5} \hat{\Psi}_\mathbf{x}(t))  \right]\right\rangle\;, \nonumber
\end{align}
one also finds an approximate conservation law for the net chiral charge density of the system, such that the sum of the chiral charge density of fermions 
\begin{align}
n_5(t)=N_{f}\int_{V} \left\langle \frac{1}{2} [\hat{\bar{\Psi}}_\mathbf{x}(t), \gamma^0\gamma^5 \hat{\Psi}_\mathbf{x}(t)]\right\rangle
\end{align}
and magnetic helicity 
\begin{align}
n_{h}(t)= \frac{e^2}{4\pi^2}\int_0^t dt' \int_{V}~\mathbf{E}_\mathbf{x}(t')\cdot \mathbf{B}_\mathbf{x}(t')
\end{align}
is conserved in the chiral limit ($m\to0$) and we have checked explicitly that for the small values of $m$ considered, dissipative effects due to finite fermion mass in the second line of Eq.~(\ref{eq:anomalybudget}) are negligible over the time scale of our simulations. 

\begin{figure}
\begin{center}
\includegraphics[width=0.5\textwidth]{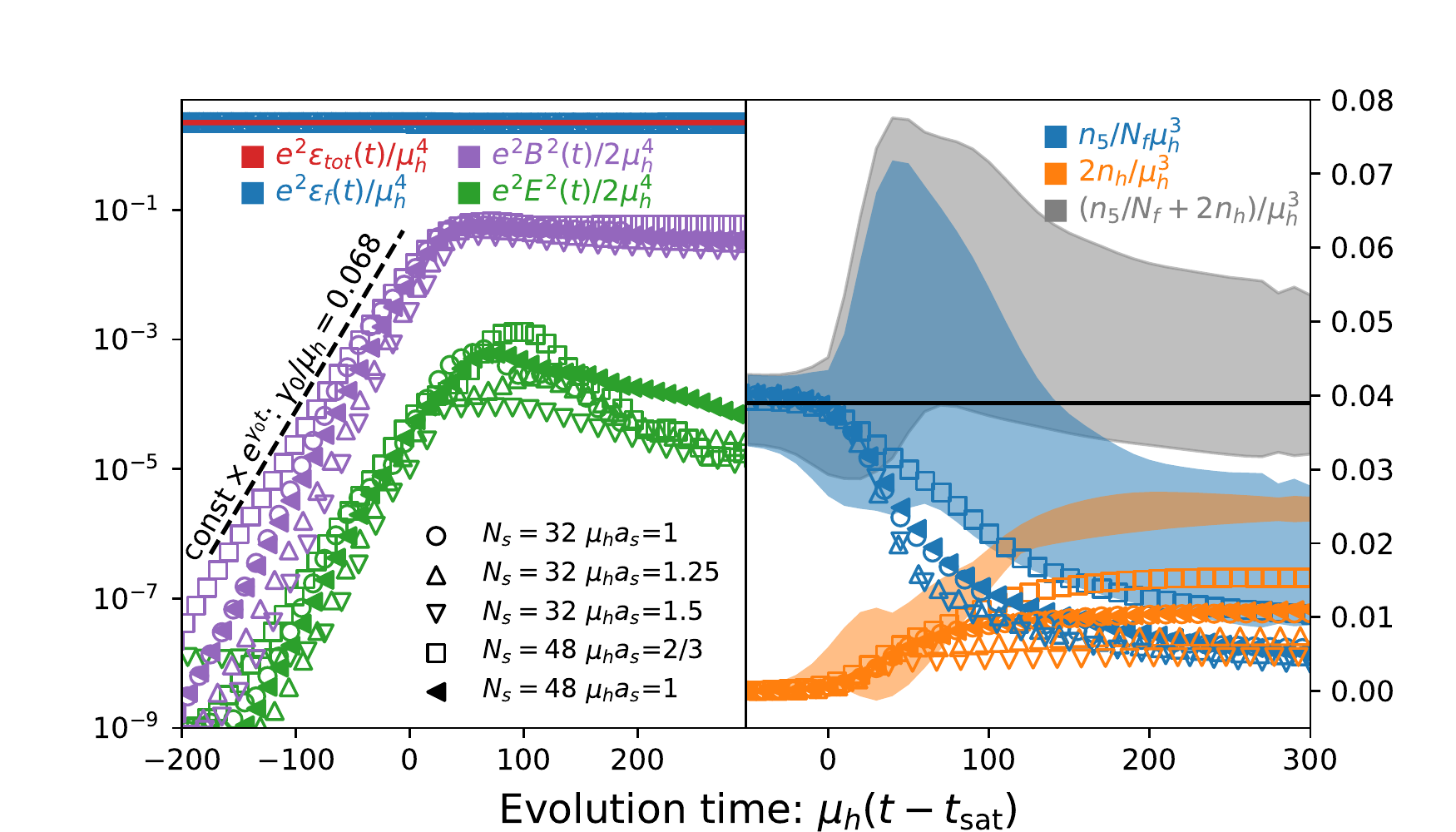}
\end{center}
\caption{Evolution of the individual contributions to the energy density $\varepsilon_{\rm tot}(t)$ (left) and chiral charge density $n_5(t)/N_f+2 n_h(t)$ (right).  Different symbols correspond to results obtained using different discretizations $\mu_ha_s=0.66 - 1.5$ on $N_s^3=32^3,48^3$ lattices. Shaded bands in the right panel show continuum extrapolations (see Supplement \cite{SM}) which satisfy the anomaly relation in Eq.~(\ref{eq:anomalybudget}) to good accuracy.}
\label{fig:anomalyplot}
\end{figure}

Simulation results for the individual contributions to the energy density (left panel) and net chirality (right panel) are compactly summarized in in Fig.~\ref{fig:anomalyplot}. Different points in each panel show the results for different lattice sizes and spacings, and we have shifted the horizontal axis of the individual data sets to account for the residual discretization dependence in the time $\mu_h \tsubsat$ where exponential growth saturates~\footnote{Saturation times $\tsubsat$ vary between $240$ for $\mu_{h}a_s=1.5$ and $375$ for $\mu_{h}a_{s}=0.66$ (see \cite{SM}), but are approximately independent of the lattice volume.}. While initially the dominant contribution to energy density and net chirality resides in the fermion sector, the chiral plasma instability leads to an exponential growth of electric and magnetic components of the energy density. Growth rates of volume averaged quantities $\mathbf{E}^2(t)$,~$\mathbf{B}^2(t)$ and $n_{h}(t)$ are dominated by the growth rate $\gamma_0$ of the maximally unstable mode as indicated by the dashed line $\propto e^{\gamma_0 t}$. Despite the exponential increase, only a small fraction of the total energy density $e^2\varepsilon_{\rm tot} \approx 0.033 ~e^2N_f~\mu_h^4$ is transferred from fermions to electromagnetic fields. It is also interesting to observe that throughout the evolution, the magnetic field strength exceeds the electric one by at least one order of magnitude, $\mathbf{B}^2 \gg \mathbf{E}^2$, indicating the presence of strong interactions between gauge and matter fields. 

When considering the balance of the net chirality in the plasma, a manifestly different picture emerges. Despite the fact that the (continuum) anomaly relation in Eq.~(\ref{eq:anomalybudget}) is violated at finite lattice spacing, our use of operator improvements significantly reduces discretization effects allowing us to perform controlled continuum extrapolations, which are consistent with conservation of the net chirality, as is indicated by the shaded bands in the right panel of Fig.~\ref{fig:anomalyplot}. While the continuum extrapolation is subject to relatively large uncertainties due to the available lattice sizes, we can safely infer that a substantial amount of the axial charge density of fermions $n_{5}(t=0) \approx 0.039 \mu_h^3 N_f$  is transferred to magnetic helicity density over the course of the evolution. Specifically, we find that for the two largest lattices available, the magnetic helicity eventually dominates over the axial charge of fermions, i.e. $n_{h}(t) \gtrsim n_{5}(t)/N_f$ at late times.

\begin{figure}[tb]
\begin{center}
\includegraphics[width=0.47\textwidth]{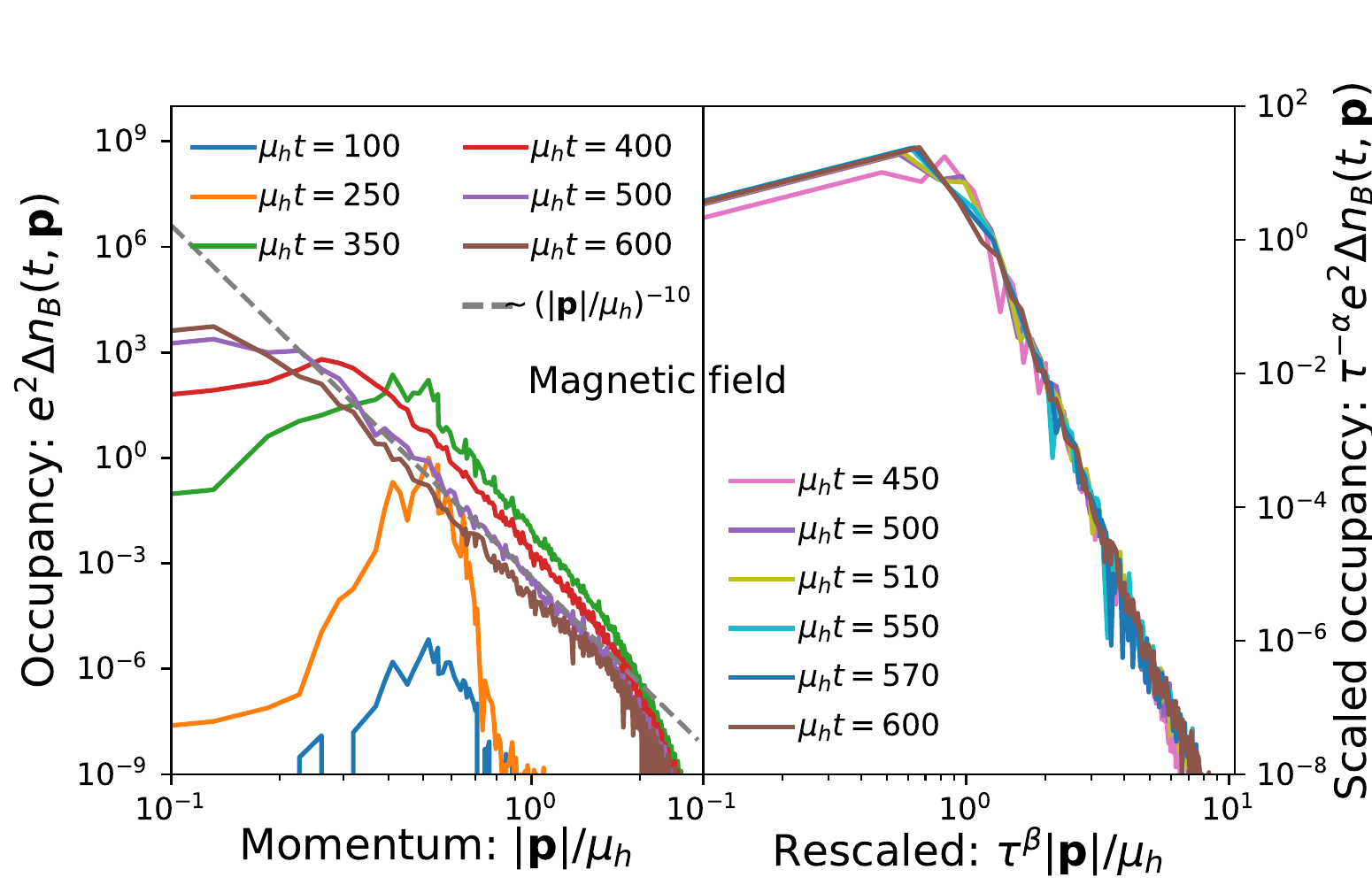} \\
\includegraphics[width=0.47\textwidth]{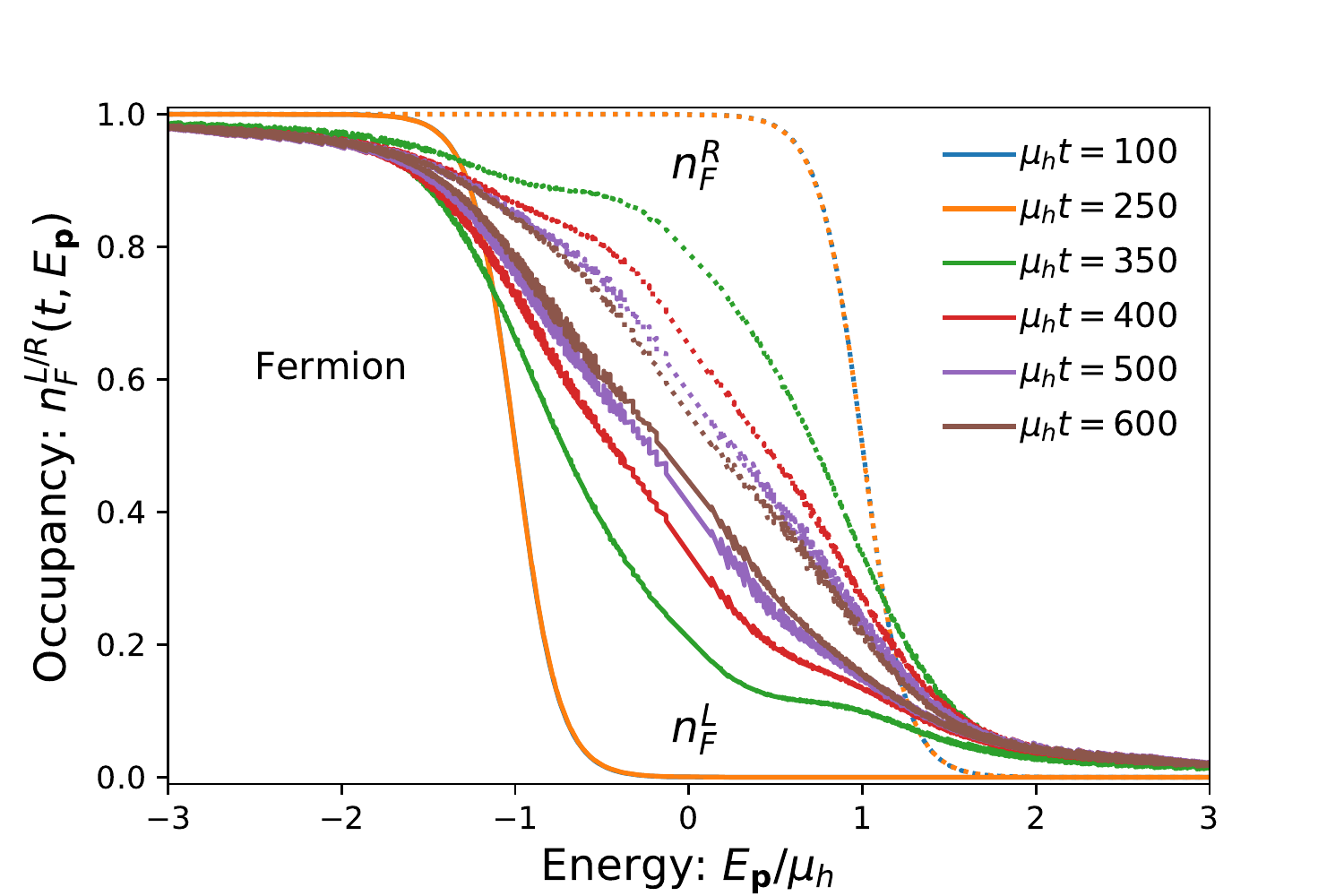}
\end{center}
\caption{(top left) Evolution of the net magnetic helicity spectrum $\Delta n_{B}(t,\mathbf{p})$ during the instability and subsequent turbulent regime. (top right) Self-similarity of the net magnetic helicity spectrum in the turbulent regime, for evolution times $\mu_h t=450, \cdots,600$ re-scaled according to Eq.~\ref{eq:Scaling} with $\alpha=0.9$,~$\beta=0.3$ and $\mu_ht^{*}=375$. (bottom) Evolution of the spectrum $n_{F}^{L/R}(t,E_{\mathbf{p}})$ of left/right handed fermions shows a clear depletion of the axial charge imbalance.}
\label{fig:gspectra}
\end{figure}

\textit{Chiral Turbulence.} 
Subsequent to the saturation of unstable growth, the plasma enters a turbulent regime characterized by a much slower evolution of the system, which we analyze in terms of the {(dimensionless)} magnetic field and fermion spectra depicted in Fig.~\ref{fig:gspectra} at different times $\mu_ht = 100-600$. Spectral distributions of the net-helicity in the gauge field sector,  i.e. the difference between occupation numbers of left- and right-handed magnetic field modes 
\begin{align}
\Delta n_{B}(t,\mathbf{p})= n^{R}_{B}(t,\mathbf{p}) - n^{L}_{B}(t,\mathbf{p})\,,
\end{align} 
are presented in the top panel, whereas the lower panel shows the spectra of left- and right-handed fermions 
\begin{align}
n_F^{L/R}(t,\pm E_{\mathbf{p}})= \left\langle   \hat{\Psi}^{\dagger}_{\mathbf{p}}(t) u_{L/R}^{\pm}(\mathbf{p})u^{\pm\dagger}_{L/R}(\mathbf{p}) \hat{\Psi}_{\mathbf{p}} (t)\right\rangle\,,
\end{align}
extracted from gauged fixed equal time correlation functions of the fermion field $\hat{\Psi}_{\mathbf{p}} (t)= \frac{1}{\sqrt{V}} \int_{\mathbf{x}} \hat{\Psi}_\mathbf{x}(t)~e^{-i\mathbf{p}\cdot\mathbf{x}}$ in Coulomb gauge by performing the appropriate projections onto left- and right-handed helicity spinors $u_{L/R}^{\pm}(\mathbf{p})$ of particles (+) and anti-particles (-).

 Starting from the linear instability regime for $\mu_{h} t \lesssim 250$, the net-helicity in the gauge sector shows an exponential growth within a limited range of wave numbers $|\mathbf{p}| \lesssim 0.8 \mu_{h}$, while left and right handed fermion spectra remain essentially unchanged with distinct sharp Fermi surfaces separated by the helicity chemical potential, shown in the lower panel of \Fig{fig:gspectra}. Secondary growth of instabilities between $\mu_{h} t \approx 250-300$ leads to a strong population of magnetic field modes at low and high wave-numbers. Over the same period of time the rapid changes in the gauge field sector are accompanied by a significant heating and depletion of the helicity imbalance in the fermion sector, as can be inferred from the softening of the Fermi surface along with narrowing of the gap between left and right handed modes. Eventually, for $\mu_ht \gtrsim 300$, the growth of the chiral instability saturates, and the evolution slows down considerably compared to the rapid changes at earlier times.

In the turbulent regime the spectrum of magnetic helicity $\Delta n_{B}(t,\mathbf{p})$ exhibits a self-similar scaling behavior, which is illustrated in the top right panel of Fig.~\ref{fig:gspectra}. Upon re-scaling, the spectra at different evolution times $\mu_{h} t = 450 -600$ are all found to collapse onto a single scaling curve $f_s(|\mathbf{p}|)$.  While a detailed characterization of the scaling function $f_s(|\mathbf{p}|)$ is beyond the scope of this work, we note that for intermediate momenta, the scaling function $f_s(|\mathbf{p}|)\sim |\mathbf{p}|^{-\kappa}$ features a  power law behavior with a large scaling exponent $\kappa = 10.2 \pm 0.5$ illustrated by the gray dashed line. Due to self-similarity, the late time evolution of the spectrum of magnetic helicity can be characterized in a compact form
\begin{eqnarray}
\label{eq:Scaling}
e^2 \Delta n_{B}(t,\mathbf{p})=\tau^\alpha f_s(\tau^\beta|\mathbf{p}|), 
\end{eqnarray}
with scaling exponents $\alpha,\beta$ and scaling function $f_{s}$, where $\tau\equiv \mu_h(t-t^*)$ is a dimensionless time variable with respect to the reference time $\mu_ht^*\approx 375$ for the transition to the turbulent regime. Notably, a self-similiar behavior as in Eq.~(\ref{eq:Scaling}) is characteristic for the late stage evolution of unstable systems, and has been reported previously in a variety of different contexts \cite{Micha:2004bv,Schmied:2018mte,Berges:2013fga}. In all of these examples, the initial instability leads to a rapid memory loss of the initial conditions, such that the subsequent turbulent evolution is universal and entirely characterized by $\alpha,\beta$ and $f_s$, which describe the transport of a conserved quantity across a large separation of scales. Based on a statistical scaling analysis, following the procedures outlined in~\cite{Berges:2013fga} (see supplementary material \cite{SM} for details), we obtain the following estimates for the scaling exponents
\begin{align}
\alpha = 1.14\pm 0.50\,, \qquad \beta = 0.37\pm 0.13\,. \nonumber
\end{align}
Since the magnetic helicity density $n_{h}$ can be equivalently expressed as an integral over the net helicity spectrum 
$n_{h}(t)=e^2\int \frac{d^3\mathbf{p}}{(2\pi)^3} \Delta n_{B}(t,\mathbf{p}),$
 one finds that the approximate validity of the scaling relation $\alpha\approx 3 \beta$ implies the conservation of the magnetic helicity of the plasma at late times, which is consistent with the behavior seen in Fig.~\ref{fig:anomalyplot}. One therefore concludes that the self-similar behavior in Eq.~(\ref{eq:Scaling}) should be associated with an inverse cascade of magnetic helicity, which is transported from microscopic ($ \ell \sim \mu_{h}^{-1}$) to macroscopic length scales ($\ell\sim \mu_{h}^{-1} \tau^{\beta}$) {\footnote{In order to exclude the possibility that the inverse cascade may occurs due to artificial non-linearities introduced by the lattice formulation, we have verified explicitly that the inverse cascade does not occur when the gauge matter interaction is switched off at intermediate simulation times. Instead the resulting gauge field spectrum remains approximately time independent as expected for non-interacting electro-magnetic fields.}}.

\begin{figure}[tb]
\begin{center}
\includegraphics[width=0.49\textwidth]{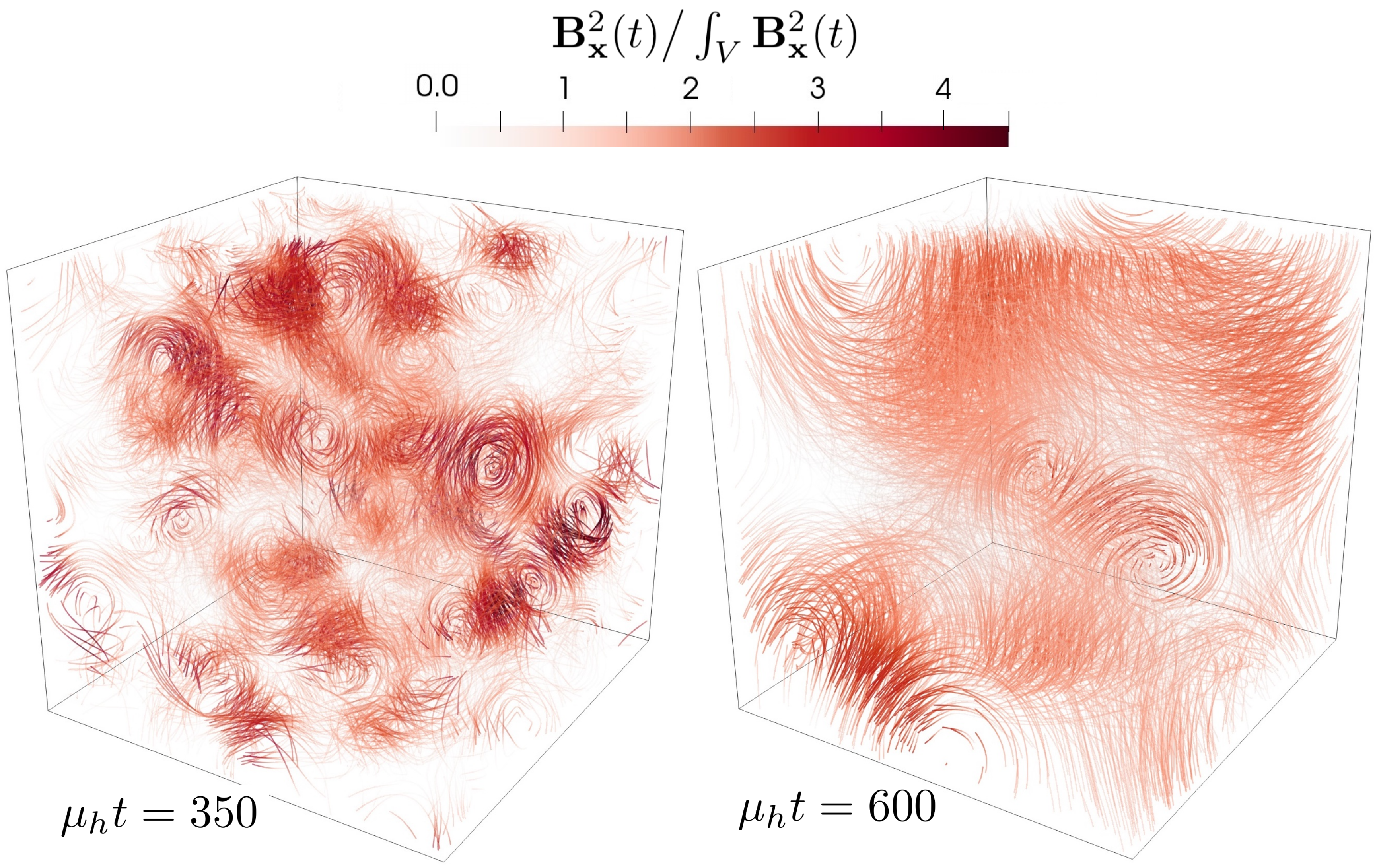}
\end{center}
\caption{ Visualization of magnetic field lines at times $\mu_ht=350$ (left) and $\mu_h t=600$ (right). Coarsening of the magnetic field lines due to the inverse cascade of magnetic helicity is observed.} 
\label{fig:3Dviz}
\end{figure}

Strikingly, the inverse cascade of magnetic helicity also manifests itself directly in the spatial structure of the magnetic field configurations, as illustrated in Fig.~\ref{fig:3Dviz} where we show stream tracing plots of the magnetic field lines, colored by the relative magnetic field intensity $\mathbf{B}_\mathbf{x}^2(t)/ \int_{V} \mathbf{B}_\mathbf{x}^2(t)$. Starting from a significant number of small scale swirls at $ \mu_h t= 350$ where the unstable growth saturates, one observes a clear coarsening of the magnetic fields towards later times $ \mu_h t=600$, where a few swirls fill the entire simulation volume. It is also evident from Fig.~\ref{fig:3Dviz} that the QED plasma develops sizeable inhomogeneities over the course of the evolution of the chiral instability, which also manifest themselves in other observables such as e.g. vector/axial charge densities which are not shown here but will be discussed in a forthcoming publication~\cite{in-prep}.

\textit{Conclusions \& Outlook.} We presented an ab-initio study of chiral instabilities and chiral turbulence, based on microscopic real-time lattice simulations of strongly coupled QED. Chirality transfer through chiral instabilities and the subsequent generation of macroscopic helical magnetic fields proceeds in a three stage sequence. Initial primary growth is followed by secondary growth until the instability saturates, when the gauge field occupation numbers become non-perturbatively large.  During the unstable phase, the fermion chirality is significantly depleted and transferred into magnetic helicity, while most of the energy is still carried by fermions. Subsequently, the system enters a turbulent regime, where magnetic helicity is transported to large distances by an inverse cascade.

While our current study established the dynamics of chiral turbulence close to the chiral limit ($m \ll \mu_{h},T$) for strongly coupled QED plasmas ($e^2 N_f \gg 1$), one important next step would be to explicitly verify the universality of our results by varying the coupling strength $e^2N_f$ and further explore the impact of dissipative effects due to finite fermion mass on the chiral turbulent regime. With regards to the dynamics of the Chiral Magnetic Effect in QCD, it would also be interesting to investigate and compare the analogous dynamics in non-Abelian gauge theories, where one ultimately expects the chirality imbalance in the fermion sector to be absorbed into a non-trivial topology of the non-Abelian gauge fields~\cite{McLerran:1990de}.

\textit{Acknowledgements.}
We gratefully acknowledge the opportunity to participate in the 2018 Intel Knights Landing Hackathon hosted by Brookhaven National Laboratory (BNL) and would like to extend our special thanks to P.~Steinbrecher for his help in optimizing the simulation software. The BNL Hackathon was partially supported by the HEP Center for Computational  Excellence (KA24001022) and the SOLLVE Exascale Computing Project  (17-SC-20-SC). BNL is supported by the Office of Science of the U.S. Department of Energy under Contract No. DE-SC0012704.
We would like to thank the Institute for Nuclear Theory at the University of Washington and the ITP Heidelberg for their kind hospitality during the completion of this work. 
M.M. was supported by the European Research Council grant ERC-2015-CoG-681707. N.M. is supported by the U.S. Department of Energy, Office of Science, Office of Nuclear Physics, under contract No. DE- SC0012704 and by the Deutsche Forschungsgemeinschaft (DFG, German Research Foundation) - Project 404640738.  Sa.S. gratefully acknowledges partial support from the Department of Science and Technology, Govt. of India through a Ramanujan fellowship and from the Institute of Mathematical Sciences. This research used resources of the National Energy Research Scientific Computing Center (NERSC), a U.S. Department of Energy Office of Science User Facility operated under Contract No. DE-AC02-05CH11231.

\bibliographystyle{apsrev4-1} 
\bibliography{references}

\section*{Supplemental material}
\subsection{Details of lattice implementation}
We use the compact Hamiltonian formulation of $U(1)$ gauge theory (in the temporal gauge) discretized on a  three dimensional $N_s\times N_s \times N_s$ lattice with periodic boundary conditions and spacing $a_s$ along each of the three spatial directions. Dimensionless lattice gauge links $U_{i}$ and electric field variables $E_{i}=ea_s^2 E_{i}^{\rm cont}$ are defined at each spatial lattice site $\mathbf x$ and each instant of time $t$ as,
\begin{eqnarray}
\label{eq:UEvo}
U_{i}(t,\mathbf{x}) &=& \exp (-ie a_{s} A_{i}(t,\mathbf{x})  )\;, \\
\label{eq:EEvo}
E_{i}(t,\mathbf{x}) &=& {e a_s^2}~\partial_t A_{i}(t,\mathbf{x})\;,
\end{eqnarray}
where $e$ is the electromagnetic coupling. Similarly we define a dimensionless fermion field operator $\hat{\Psi}(t,\mathbf{x})=a_s^{3/2}\hat{\Psi}^{\rm cont}(t,\mathbf{x}) $, which is discretized by performing a mode function decomposition in the discrete momentum $\mathbf q$ and helicity $h=\pm 1$ basis. Denoting the sum over all $N_s^3$ lattice momenta as $\sum_{\mathbf{q}}$ the mode function decomposition of the fermion operator then takes the form
\begin{align}
\label{eq:ModeDecomposition}
\hat{\Psi}(t,\mathbf{x}) = \frac{1}{\sqrt{N_{s}^3}}\sum\limits_{{\s}} \, \Big[\Phi^+_{\s}(t,\mathbf{x}) \, \hat{b}_{\s}+\Phi^-_{\s}(t,\mathbf{x}) \, \hat{d}^\dagger_{\s}\Big]\,,
\end{align}
where $ \Phi^{\pm}_{\s}(t,\mathbf{x})$ are time dependent wave functions for particles and anti-particles respectively.  Vice versa  $\hat{b}_\s=b_{\s}^{\rm cont}/\sqrt{N_s^{3}a_s^{3}}$ and $\hat{d}_\s^\dagger$ are dimensionless and  time independent creation and annihilation operators of fermions acting on the initial state, such that $b^{\dagger}_{\s}/d^{\dagger}_{\s}$ creates a particle/anti-particle with momentum $\mathbf{q}$, helicity $\pm h$ and electric charge $\pm e$ in the initial state.

Based on the above definitions of lattice variables, the discretized version of the QED Hamiltonian $\hat{H}$ takes the form
\begin{align}
&e^2 a_s \hat{H} = e^2 N_f \sum_{\mathbf{x}} \frac{1 }{2}  \left[\hat{\Psi}^{\dagger}(t,\mathbf{x}), \gamma^0 \big(   -ia_s\slashed{D}^W_s +ma_s  \big) \hat{\Psi}(t,\mathbf{x})\right]
  \nonumber\\
  & \quad + \sum\limits_{\mathbf{x},i}  \frac{E^2_i(t,\mathbf{x})}{2}+\sum_{\mathbf{x},i< j}
  \left[ 1- \frac{1}{2} \left( U_{i,j}(t,\mathbf{x})  + U_{i,j}^{\dagger}(t,\mathbf{x})\right) \right]
\end{align}
where $\sum_{i}$ denotes the sum over $x,y,z$ directions and $\sum_\mathbf{x}$ denotes the sum over all $N_s^3$ lattice sites.  Denoting the unit lattice vectors in the $i,j$ direction as $\mathbf{i},\mathbf{j}$, the magnetic contribution to the lattice Hamiltonian is given in terms of the gauge invariant spatial plaquette $U_{i,j}(t,\mathbf{x})$ which is defined as usual as
\begin{eqnarray}
\label{eq:Plaquette}
U_{i,j}(t,\mathbf{x}) = U_i(t,\mathbf{x}) U_j(t,\mathbf{x}+\mathbf{i}) U^\dagger_i(t,\mathbf{x}+\mathbf{j}) U^\dagger_j(t,\mathbf{x})\;.
\end{eqnarray}
With regards to the fermion discretization, we employ an $\mathcal O(a_s^3)$ tree-level improved Wilson-Dirac operator
 \begin{align}
-i&a_s\slashed{D}^W_s \hat{\Psi}(t,\mathbf{x})
\nonumber\\
&=   \frac{1}{2} \sum\limits_{n,i} C_n \left[ \left(-i\gamma^i-nr_W\right) U_{+ni}(t,\mathbf{x}) \hat{\Psi}(t,\mathbf{x}+n\hat{\mathbf{i}}) \right.
\nonumber\\
&+\left. 2n r_W \hat{\Psi}({\mathbf{x}},t) - \left(-i\gamma^i + nr_W\right) U_{-ni} (t,\mathbf{x}) \hat{\Psi}(t,\mathbf{x}-n\hat{\mathbf{i}}) \right]\,,
 \end{align}
with the coefficients $C_{1}=4/3$, $C_{2}=-1/6$ and $C_{n> 2}=0$, and the Wilson parameter is chosen to be $r_W=1$. We note that the $n$-link terms in the Dirac operator involve the product of multiple forward and backward going $U(1)$ gauge links, which are explicitly given by 
\begin{align}
&U_{+ni}(t,\mathbf{x})= \prod_{k=0}^{n-1} U_{i}(t,\mathbf{x}+k \mathbf{i})\;, 
\nonumber\\
&U_{-ni}(t,\mathbf{x})= \prod_{k=1}^{n} U^{\dagger}_{i}(t,\mathbf{x}-k \mathbf{i})\;.
\end{align}
Discretized lattice gauge links $U_{i}(t,\mathbf{x})$, electric fields $E_{i}(t,\mathbf{x})$ and fermion wave functions $\Phi^{\pm}_{\s}(t,\mathbf{x})$ are then evolved as a function time using the equations of motion derived from the lattice discretized QED Hamiltonian~\cite{Mace:2016shq}. 
\subsubsection{Initial Conditions}\label{sec:init}
Before we comment on the details of this procedure, we briefly discuss the implementation of the initial conditions for fermions and electro-magnetic fields. Within the classical-statistical approach the initial conditions for the fermion sector are specified in terms of the expectation values of all relevant products of creation and annihilation operators $\hat{b}_\s$ and $\hat{d}_\s^\dagger$ in the initial state along with the initial conditions for the corresponding wave-function $\Phi^{\pm}_{\s}(t,\mathbf{x})$. By working in the discrete momentum $\mathbf q$ and helicity $h$ basis, we specify the relevant operator expectation values in terms of the initial occupation numbers of particles and anti-particles
\begin{eqnarray}
\label{eq:OPEpectations}
\frac{1}{2}\left\langle \Big[ \hat{b}^{\dagger}_{\s},\hat{b}_{\sprime} \Big] \right \rangle&=&+\Big[n^+_{\s}(t=0)-\frac{1}{2}\Big] \delta_{\mathbf{q}\mathbf{q}'}\delta_{hh'}\;, \nonumber \\
\frac{1}{2}\left\langle\Big[ \hat{d}_{\s},\hat{d}^{\dagger}_{\sprime} \Big]\right\rangle&=&-\Big[n^-_{\s}(t=0)-\frac{1}{2}\Big] \delta_{\mathbf{q}\mathbf{q}'}\delta_{hh'}\;, \nonumber \\
\end{eqnarray}
which in accordance with our discussion in the main text are given by a Fermi-Dirac distribution at inverse temperature $\beta$ and helicity chemical potential $\mu_{h}$ as 
\begin{eqnarray}\label{eq:DefFermiDiracDistT0}
n^{\pm}_{\s}(t=0)= \frac{1}{\rm{e}^{ \beta( |E_\mathbf{p}| \mp h\mu_h) } +1}\,.
\end{eqnarray}
By inserting the operator decomposition in Eq.~(\ref{eq:ModeDecomposition}), the expectation values of a generic operator which is bi-linear in the fermion fields can be evaluated according to
\begin{align}\label{eq:fermionblinear}
&\langle \hat{\Psi}^{\dagger}(t,{\mathbf{x}}) O(t,{\mathbf{x},\mathbf{y}} )\hat{\Psi}(t,{\mathbf{y}}) \rangle 
\nonumber\\
&=  \frac{1}{N_s^3} \sum_{\s} \Big\{\Phi_{\s}^{+\dagger}(t,\mathbf{x}) O(t,{\mathbf{x},\mathbf{y}} ) \Phi^+_{\s}(t,\mathbf{y})\Big[n^+_{\s}(t=0)-\frac{1}{2}\Big]
\nonumber\\
&
-\Phi_{\s}^{-\dagger}(t,\mathbf{x})  O(t,{\mathbf{x},\mathbf{y}} )\Phi^-_{\s}(t,\mathbf{y})\Big[n^-_{\s}(t=0)-\frac{1}{2}\Big] \Big\}\,.
\end{align}
where $O(t,{\mathbf{x},\mathbf{y}})$ is the lattice discretized version of the respective observable.

While the operator expectation values in Eq.~(\ref{eq:OPEpectations}) describe the physical state of the system at initial time, the initial conditions for the fermion wave functions $\Phi^{\pm}_{\s}(t,\mathbf{x})$ in the discrete momentum $\mathbf q$ and helicity $h$ basis are simply given in terms of plain wave spinors with definite helicity 
\begin{eqnarray}\label{eq:defmodeswavefunctions}
&\Phi^{\pm}_{\s}(t=0,\mathbf{x})=e^{\pm \frac{2\pi i \, \mathbf{q}\cdot\mathbf{x}}{N_s}} u^{\pm}_{h}(\mathbf{p})\,.
\end{eqnarray}
where $u^\pm_{h}(\mathbf{p})$ denote the helicity spinors for Wilson-Dirac fermions
\begin{eqnarray}
\frac{\mathbf{p} \cdot \mathbf{\Sigma}}{|\mathbf{p}|}~u^\pm_{h}(\mathbf{p}) = \pm h u^\pm_{h}(\mathbf{p})\;,
\end{eqnarray}
with normalization
\begin{eqnarray}
u^{\pm \dagger}_{h}(\mathbf{p}) u^\pm_{h'}(\mathbf{p})&=&\delta_{hh'}\;, \\ u^{\pm\dagger}_{h}(\mathbf{p}) u^\mp_{h'}(-\mathbf{p})&=&0\;.
\end{eqnarray}
Explicitly we construct the helicity spinors in the Dirac represenation as
\begin{align}\label{eq:modefunctions}
&u^\pm_h(\mathbf{p})=\frac{|\mathbf{p}|}{\sqrt{2 E_{\mathbf{p}}(E_{\mathbf{p}} - m_{W})}}\begin{pmatrix}
\varphi_h(\pm\mathbf{p})\\
\pm h\frac{E_{\mathbf{p}}- m_{W}}{|\mathbf{p}|}\varphi_h(\pm\mathbf{p})
\end{pmatrix}\,.
\end{align}
where $E_{\mathbf{p}}= \pm \sqrt{\mathbf{p}^2+m_W^2}$ is the energy eigenvalue, with $\mathbf{p}$ and $m_{W}$ denoting the momenta and effective masses of the $\mathcal O(a_s^3)$ improved Wilson Dirac fermions given by
\begin{eqnarray}
{p}_i&=& \sum_{n} \frac{C_n}{a_s} \sin\left( \frac{ 2\pi n {q}_i }{N_s}\right)\;, \\
m_{W}&=& m + 2r_W\sum_{i,n} \frac{C_n}{a_s}\sin^2\left( \frac{\pi  n {q}_i}{N_s}\right)\;,
\end{eqnarray}
and $\varphi_h(\mathbf{p})$ are two component Weyl spinors 
\begin{eqnarray}
\frac{\mathbf{p} \cdot \boldsymbol{\sigma}}{|\mathbf{p}|}  \varphi_h(\mathbf{p}) = h \varphi_{j}(\mathbf{p})
\end{eqnarray}
normalized to unity $\varphi_h^{\dagger} (\mathbf{p}) \varphi_{h'}(\mathbf{p}) =\delta_{hh'}$. Based on the above definitions, it is straightforward to verify that the initial state of the system 
features a non-vanishing axial charge density
\begin{eqnarray}
a_s^3 n_{5}(t=0) =   \frac{N_f}{N_s^3} \sum_{\mathbf{x}} \frac{1}{2}\Big \langle  \left[\Psi^{\dagger}(t=0,\mathbf{x}), \gamma_5 \Psi(t=0,\mathbf{x}))\right] \Big\rangle\;, \nonumber
\end{eqnarray}
which is explicitly given by
\begin{eqnarray}
a_s^3 n_{5}(t=0)=\frac{N_f}{N_s^3} \sum_{\s}  \frac{h|\mathbf{p}|}{|E_{\mathbf{p}}|} \left( n^{+}_{\s}(t=0) - n^{-}_{\s}(t=0)\right)\;. \nonumber
\end{eqnarray}

We initialize the lattice gauge links and electric fields at initial time $t=0$ as a random superposition of plane waves to describe their initial vacuum fluctuations~\cite{Berges:2013fga}. Denoting the two transverse polarization vectors as $\boldsymbol{\epsilon}_{\lambda}(\mathbf q)$ with $\lambda=1,2$ as described in \cite{Berges:2013fga,Mace:2016svc}, the Fourier components of the initial gauge fields are then given by
\begin{align}
\label{eq:ICSGaugeFourier}
e\tilde{A}_i(0,\mathbf{q}) &= \frac{1}{\sqrt{2|\tilde{\mathbf{p}}|a_s}} \sum_\lambda\,\Big[ c_{\lambda}(\mathbf{q}) \boldsymbol{\epsilon}^{\lambda}_{i}(\mathbf{q})  + c^*_{\lambda} (-\mathbf{q})  \boldsymbol{\epsilon}^{\lambda*}_{i}(-\mathbf{q}) \Big]\,, 
\nonumber\\
\tilde{E}_i(0,\mathbf{q}) &= i\sqrt{ \frac{|\tilde{\mathbf{p}}|a_s}{2}} \sum_\lambda\,\Big[ c_{\lambda}(\mathbf{q}) \boldsymbol{\epsilon}^{\lambda}_{i}(\mathbf{q})  - c^*_{\lambda} (-\mathbf{q})  \boldsymbol{\epsilon}^{\lambda*}_{i}(-\mathbf{q}) \Big]\,.
\end{align}
where $|\tilde{\mathbf{p}}|$ is the dispersion momentum of the lattice mode
\begin{eqnarray}
\label{eq:latMomBos}
|\tilde{\mathbf{p}}|=\sqrt{\tilde{{p}}_{i} \tilde{{p}}^{*}_{i}}\;, \qquad \tilde{{p}}_{j}=-i\frac{1-\rm{e}^{-\frac{2\pi i  {q}_{j}}{N_s}}}{a_s}\;,
\end{eqnarray}
 and $c_\lambda(\mathbf{q})$ are complex random numbers distributed according to Gaussian distribution, with $\langle c_\lambda(\mathbf{q}) \rangle=0$ and
\begin{eqnarray}
\label{eq:cStat}
\left\langle c_\lambda^*(\mathbf{q}) c_{\lambda'}(\mathbf{q}') \right\rangle = \frac{e^2}{2}~\delta_{\mathbf{q}\mathbf{q}'}\delta_{\lambda\lambda'} ,
\end{eqnarray}
to describe the initial vacuum fluctuations. While the electromagnetic coupling $e$ scales out of the classical equations of motion, it explicitly enters in the initial conditions for the gauge fields in Eq.~(\ref{eq:cStat}). However, since the instability results in an exponential growth of the initial gauge fields, one only expects a logarithmic sensitivity of the results to the coupling strength $e^2$ \cite{Berges:2013fga} and we have therefore chosen a relatively small value of $e^2/2 = 0.5 \cdot 10^{-6}$ to realize a sufficiently long period of exponential growth to reliably extract growth rates.

Based on Eq.~(\ref{eq:ICSGaugeFourier}), the initial conditions for the lattice gauge links and electric fields in coordinate space are then obtained via a Fourier transformation
\begin{align}
&ea_s{A}_i(0,\mathbf{x}) = \frac{1}{\sqrt{N_s^3}} \sum_{\mathbf{q}}\,  e\tilde{A}_i(0,\mathbf{q})\; \rm{e}^{\frac{  {2\pi i \mathbf{q}\cdot \mathbf{x} }}{{N_s}} }  \,, 
\nonumber\\
&{E}_i(0,\mathbf{x}) = \frac{1}{\sqrt{N_s^3}} \sum_{\mathbf{q}}\, \tilde{E}_i(0,\mathbf{q})\; \rm{e}^{\frac{  {2\pi i \, \mathbf{q}\cdot \mathbf{x}}}{{N_s}} }  ~.
\end{align}
with subsequent exponentiation of $ea_s{A}_i(0,\mathbf{x})$ according to Eq.~(\ref{eq:UEvo}) to obtain the initial gauge links $U_i(0,\mathbf{x})$.

\subsubsection{Hamiltonian time evolution scheme}
\begin{figure}[t!]
\centering
\includegraphics[width=0.45\textwidth]{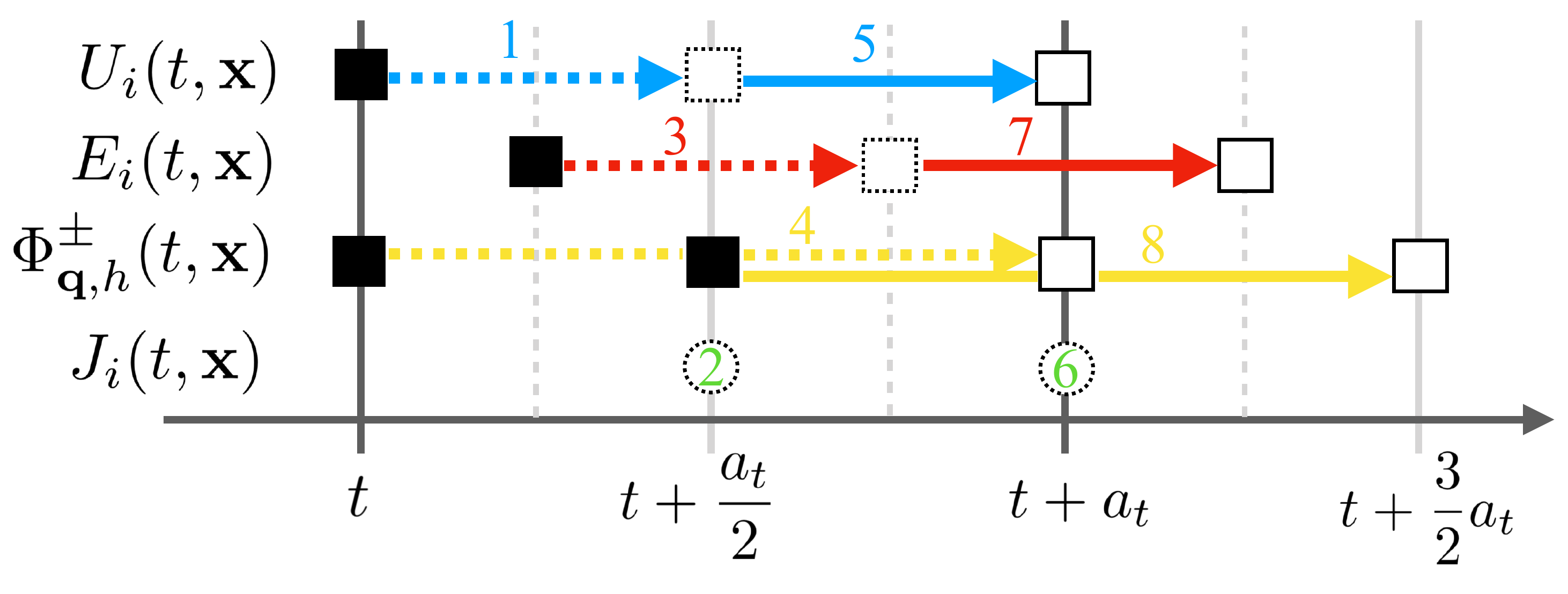}
\caption{Schematic representation of the leap-frog scheme for Hamiltonian time evolution scheme employed in this study.}
\label{fig:evolutioncartoon}
\end{figure}

We solve the time evolution of gauge links and the fermion fields numerically using a variant of a leap-frog algorithm to 
preserve time-reversal invariance. When updating the fields at each time-step, we use the gauge links $U_{i}(t,\mathbf{x})$, electric fields $E_{i}(t+a_t/4,\mathbf{x})$ as well as two sets of fermion mode functions $\Phi^{\pm}_{\s}(t,\mathbf{x})$ and  $\Phi^{\pm}_{\s}(t+a_t/2,\mathbf{x})$ defined at different instances of time. Starting from the initial conditions specified in the previous paragraph, we use the following procedure, graphically represented in Fig.~\ref{fig:evolutioncartoon}, to evolve all dynamical fields by one full time step~$a_t$. By successively repeating all of the steps below, the time evolution is then calculated up to a maximum time of interest. Defining
\begin{eqnarray}
\delta t \equiv \frac{a_t}{2}\;,
\end{eqnarray}
\begin{enumerate}
\item We evolve the links by half a time step, from $t$ to $t+\delta t$, via 
\begin{eqnarray}\label{eq:updateschemeLinks}
U_{i}(t+\delta t,\mathbf{x}) &=& e^{-i\frac{\delta t }{a_s} E_{i}(t+\delta t/2,\mathbf{x})  }U_{i}(t,\mathbf{x}) \,,
\end{eqnarray} 
 with the gauge links on the r.h.s. evaluated at $t$ and the electric fields evaluated at $t+\frac{\delta t}{2}$.
\item Next, to prepare for the update of the electric fields, we use the second set of fermion wave functions $\Phi^{\pm}_{\s}(t+\delta t,\mathbf{x})$ and the gauge links $U_{i}(t+\delta t,\mathbf{x})$ to evaluate the expectation value of the fermion vector current $J_{i}(t+\delta t,\mathbf{x})$, which is given by \cite{Mace:2016shq}
\begin{align}
\label{eq:BRCurrents}
\quad J_{i}(t+\delta t,\mathbf{x})= \sum_{n} C_{n} \sum_{k=0}^{n-1}  J_{i}^{(n)}(t+\delta t,\mathbf{x}-k\mathbf{i})\,,
\end{align}
where $J_{i}^{(n)}(t+\delta t,\mathbf{x})$ denotes the individual contributions from $n$-link terms
\begin{align}
 J_{i}^{(n)}(&t+\delta t,\mathbf{x})= \frac{1}{2} \Big\langle \Big[ \hat{\Psi}^{\dagger}(t+\delta t,\mathbf{x}),~\gamma^{0}~\Big(\gamma_{i} -i n r_W\Big) 
\nonumber\\&\times
U_{+ni}(t+\delta t,\mathbf{x})~\hat{\Psi}(t+\delta t,\mathbf{x}+n\mathbf{i}) \Big]  + h.c.\Big\rangle 
\end{align}

\item Subsequently, the electric field is evolved by half a time step $\delta t$ from $t+\delta t/2$ to $t+3\delta t/2$, based on the update rule
\begin{align}\label{eq:updatescheme}
\qquad\; &E_{i}(t+3\frac{\delta t}{2},\mathbf{x})-E_{i}(t+\frac{\delta t}{2},\mathbf{x})= \frac{e^2 N_f \delta t}{a_s} J_{i}(t+\delta t,\mathbf{x})
\nonumber\\&
-2 \sum\limits_{j\neq i}  \frac{\delta t}{a_s} \text{Im} \Big( U_{i,j}(t+\delta t,\mathbf{x}) - U_{i,-j}(t+\delta t,\mathbf{x})  \Big)\;, \nonumber\\
\end{align}
where $U_{i,j}(t+\delta t,\mathbf{x})$ are the spatial plaquette terms (c.f. Eq.~(\ref{eq:Plaquette})) and $J_{i}(t+\delta t,\mathbf{x})$ is the expectation value of the fermion vector current in Eq.~(\ref{eq:BRCurrents}).
\item Next we update the complete first set of $4N_s^3$ fermion wave functions $\Phi^{\pm}_{\s}(t,\mathbf{x})$, by calculating the action of the Wilson-Dirac Hamiltonian at $t+\delta t$ on the second set of wave-functions at $\Phi^{\pm}_{\s}(t+\delta t,\mathbf{x})$, which gives rise to the following update rule 
\begin{eqnarray}
\label{eq:PhiEvo}
&\Phi^{\pm}_{\s}(t+2\delta t,\mathbf{x}) - \Phi^{\pm}_{\s}(t,\mathbf{x}) = \\
& \qquad -i \frac{2\delta t}{a_s} \gamma^{0} \left( -i a_s \slashed{D}_{s}^{W} +ma_s  \right) \Phi^{\pm}_{\s}(t+\delta t,\mathbf{x}) \nonumber
\end{eqnarray}

\item We then repeat the entire update sequence outlined above but now updating the gauge links from $t+\delta t$ to $t + 2\delta t$, using \Eq{eq:updateschemeLinks} with the gauge links at $U_{i}(t+\delta t,\mathbf{x})$  and electric fields $E_{i}(t+3\frac{\delta t}{2},\mathbf{x})$ as input on the right-hand side.
\item Subsequently, the vector current $J_{i}(t+2\delta t,\mathbf{x})$ is calculated based on \Eq{eq:BRCurrents}, but now with the updated first set of fermion fields $\Phi^{\pm}_{\s}(t+2\delta t,\mathbf{x})$ and gauge links $U_{i}(t+2\delta t,\mathbf{x})$ as input on the right hand side.
\item Based on the vector currents $J_{i}(t+2\delta t,\mathbf{x})$ and gauge links $U_{i}(t+2\delta t,\mathbf{x})$, the electric field at $E_{i}(t+\frac{5}{2}\delta_t)$ is computed from \Eq{eq:updatescheme}
\item One update step is completed by computing the update of the second set of fermion mode functions, i.e. by evolving  $\Phi^{\pm}_{\s}(t+\delta t,\mathbf{x})$ to $\Phi^{\pm}_{\s}(t+3\delta t,\mathbf{x})$ using \Eq{eq:PhiEvo}) with the gauge links $U_{i}(t+2\delta t,\mathbf{x})$ and the first set of fermion fields $\Phi^{\pm}_{\s}(t+2\delta t,\mathbf{x})$ as input on the rhs.
\end{enumerate}
We note that the above update procedure for gauge links, electric fields and fermion fields automatically leads to the conservation of the Gauss law,
\begin{align}
\label{eqn:GaussL}
G(t,\mathbf{x})= \sum_{i} &\big[E_{i}(t,\mathbf{x}) - E_{i}(t,\mathbf{x-i}) \Big]+e^2 N_f J_{0}(t,\mathbf{x})\;,
\end{align}
where $J_0=a_s^{3} J_0^{\rm cont}$ denotes the (dimensionless) vector charge density on the lattice
\begin{eqnarray}
J_0(t,\mathbf{x}) = \frac{1}{2}\langle  [\hat{\Psi}^\dagger(t,\mathbf{x}) , \hat{\Psi}(t,\mathbf{x})]\rangle
\end{eqnarray}
which is evaluated according to \Eq{eq:fermionblinear}, i.e.
\begin{align}
J_0(t,\mathbf{x}) &=  \frac{1}{N_s^3} \sum_{\s} \Big\{\Phi_{\s}^{+\dagger}(t,\mathbf{x}) \Phi^+_{\s}(t,\mathbf{x})\Big[n^+_{\s}(t=0)-\frac{1}{2}\Big]
\nonumber\\
&-\Phi_{\s}^{-\dagger}(t,\mathbf{x})\Phi^-_{\s}(t,\mathbf{x})\Big[n^-_{\s}(t=0)-\frac{1}{2}\Big] \Big\}\,,
\end{align}
Specifically, the update procedure outlined above explicitly ensures that $G(t+a_t,\mathbf{x})=G(t,\mathbf{x})$, such that the Gauss law constraint $G(t,\mathbf{x})=0$ is automatically satisfied for the entire evolution as long as it holds for the initial condition. While the initial conditions outlined above satisfy Gauss law in the continuum, non-linear effects due to the compact $U(1)$ lattice discretization will typically lead to a small Gauss law violation of the initial conditions. In order to eliminate this problem, we follow previous works in the context of non-Abelian gauge theories and explicitly restore Gauss law at the initial time by projecting the electric fields on the constraint surface using the algorithm in Ref.~\cite{Moore:1996wn}. We have checked explicitly that violations of Gauss law stay at the level of machine precision over the course of the entire simulation.

\subsubsection{Bulk observables}
Based on the solution of the gauge links, electric fields and fermion wave-functions, bulk observables such as the individual components of the energy density or vector/axial currents, can directly be calculated from a gauge invariant operator definition (see e.g. \cite{Mace:2016shq}). Specifically, the volume averaged energy density of fermions on the lattice is given by
\begin{align}\label{eq:energydensF}
&e^2a_s^{4} \varepsilon_f(t) =\\
& \qquad  \frac{e^2N_f}{N_s^3} \sum_{\mathbf{x}}  \frac{1}{2} \langle  [\hat{\Psi}(t,\mathbf{x}), \gamma^0 \big(   -ia_s\slashed{D}_s^W +ma_s  \big) \hat{\Psi}(t,\mathbf{x})] \rangle\,, \nonumber
\end{align}
while the volume averaged energy density of the gauge fields $\varepsilon_g(t)=\varepsilon_\mathbf{E}(t)+\varepsilon_\mathbf{B}(t) $, consists of an electric and a magnetic part, which 
are given by
 \begin{align}
&e^2 a_s^{4} \varepsilon_\mathbf{E}(t) =  \frac{1}{N_s^3} \sum\limits_{\mathbf{x},i}  \frac{E^2_i(t,\mathbf{x})}{2}\,,
\\\label{eq:energydensB}
&e^2 a_s^{4} \varepsilon_\mathbf{B}(t) =\frac{1}{N_s^3}\sum_{\mathbf{x},i< j}  \big( 1- \frac{1}{2} [ U_{i,j}(t,\mathbf{x})  + h.c.] \big)\,.
\end{align}

Similarly, the chiral charge density  of fermions $n_{5}(t)/N_{f}$ is obtained (per flavor) as 
\begin{align}
&\frac{a_s^3 n_5(t)}{N_f}  = \frac{1}{N_s^3} \sum\limits_\mathbf{x} \frac{1}{2}\left\langle \Big[\hat{\Psi}_{\mathbf{x}}^{\dagger}, \gamma_5 \hat{\Psi}_{\mathbf{x}}\Big] \right\rangle\;.
\end{align}
Since the magnetic helicity density is negligible at initial time $n_{h}(t=0) \approx 0$, the magnetic helicity $n_{h}(t)$ can be directly obtained by integrating the l.h.s. of the anomaly equation, i.e.
\begin{eqnarray}
a_s^3n_h(t)=\frac{1}{4 \pi^2 N_s^{3}} \int_{0}^{t} \frac{dt'}{a_s} \sum_{x} E_i^{\rm imp}(t',\mathbf{x}) B_i^{\rm imp}(t',\mathbf{x})\;, 
\end{eqnarray}
where $E_i^{\rm imp}(t,\mathbf{x})$ and $B_i^{\rm imp}(t,\mathbf{x})$ are $\mathcal{O}(a_s^2)$ improved operator definitions of the electric and magnetic fields, constructed from the lattice gauge links and electric fields by averaging nearest and next-to-nearest neighbor contributions as described in \cite{Moore:1996wn}. 


\subsubsection{Helicity projections \& Spectra}
\begin{figure}[t!]
\centering
\includegraphics[width=0.35\textwidth]{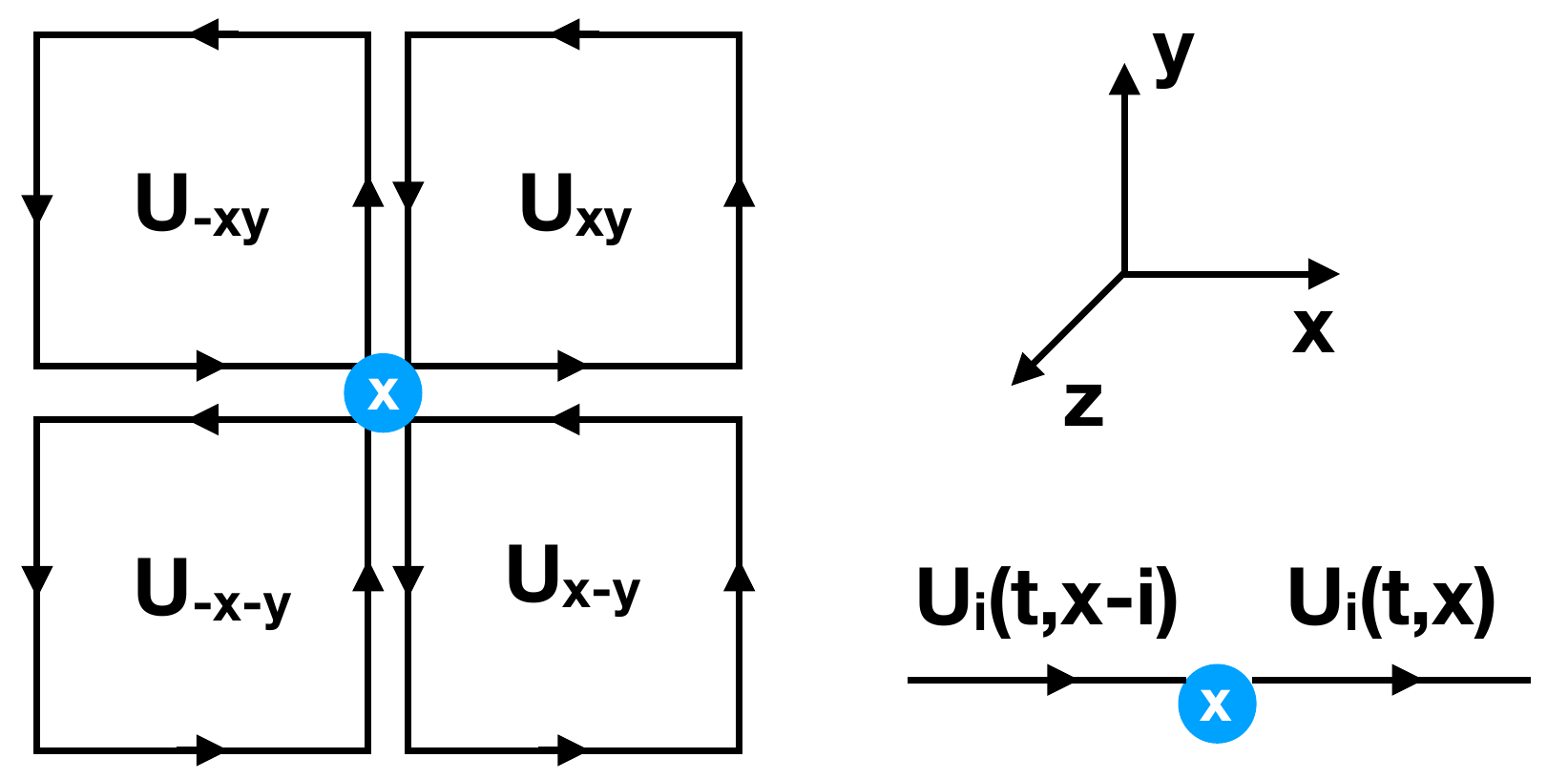}
\caption{Illustration of the operator definitions of \Eq{eq:Bdefinition} and \Eq{eq:curlBdefinition}, computed by averaging over four adjacent plaquettes
to each site.}
\label{fig:latticecoartoon}
\end{figure}

Since in abelian gauge theories the field strength tensor is gauge invariant, we can readily calculate gauge invariant spectra of electro-magnetic fields. Defining the lattice magnetic fields $\mathbf{B}(t,\mathbf{x})$ and their curl $\mathbf{\nabla} \times \mathbf{B}(t,\mathbf{x})$ at the same lattice point by averaging over neighboring plaquettes (see \Fig{fig:latticecoartoon} for an illustration of the gauge links involved in the definition)
\begin{widetext}
\begin{align}\label{eq:Bdefinition}
e a_s^2\mathbf{B}_i(t,\mathbf{x}) &= \frac{ \epsilon_{ijk}}{8}\text{Im}\, [U_{{j},{k}}(t,\mathbf{x})+U_{-{j},{k}}(t,\mathbf{x})+U_{{j},-{k}}(t,\mathbf{x})+U_{-{j},-{k}}(t,\mathbf{x})]\,,
\\
e a_s^3~\mathbf{\nabla} \times \mathbf{B}(t,\mathbf{x})&=\frac{1}{4 }
\text{Im}\, \begin{pmatrix}
U_{x,y} -U_{x-y}  -U_{z,x}+U_{-z,x} +  U_{-x,y}-U_{-x,-y} - U_{z,-x}+U_{-z,-x}  
\\
U_{y,z} -U_{y,-z}  -U_{x,y}+U_{-x,y} +  U_{-y,z}-U_{-y,-z} - U_{x,-y}+U_{-x,-y} 
\\
U_{z,x} -U_{z,-x}  -U_{y,z}+U_{-y,z} +  U_{-z,x}-U_{-z,-x} - U_{y,-z}+U_{-y,-z}
\end{pmatrix}(t,\mathbf{x})\,,\label{eq:curlBdefinition}
\end{align}
\end{widetext}
the decomposition of the magnetic fields into left/right handed components on the lattice
\begin{eqnarray}
\mathbf{B}^{L/R}(t,\mathbf{p})= \hat{P}_{L/R}(\mathbf{p}) \mathbf{B}(t,\mathbf{p}) = \frac{|\mathbf{p}|  \pm i\mathbf{p} \times }{2|\mathbf{p}|} \mathbf{B}(t,\mathbf{p})
\end{eqnarray}
can be performed in a straightforward way by recognizing $i\mathbf{p} \times \mathbf{B}(t,\mathbf{p})$ as the Fourier transform of $\mathbf{\nabla} \times \mathbf{B}(t,\mathbf{x})$ and $|\mathbf{p}|$ as the eigenvalue of the curl operator on the lattice. Based on the operator definitions in Eqns.~(\ref{eq:Bdefinition},\ref{eq:curlBdefinition}), the left- and right-handed components of the magnetic field can then be calculated as
\begin{eqnarray}
\mathbf{B}^{L/R}(t,\mathbf{p})= \hat{P}_{L/R}^{\rm lat}(\mathbf{p}) \mathbf{B}(t,\mathbf{p})
\end{eqnarray}
where $\mathbf{B}(t,\mathbf{q})$ denotes the Fourier transform of the magnetic fields defined through an elementary plaquette
\begin{eqnarray}
ea_s^{1/2}\mathbf{B}_{i}(t,\mathbf{q})=\frac{a_{s}^{3}}{\sqrt{N_s^3 a_s^3}}\sum_{\mathbf{x}} \frac{\epsilon^{ijk}}{2} \text{Im}\, [U_{{j},{k}}(t,\mathbf{x}) ] e^{-\frac{2\pi i \mathbf{q}\cdot \mathbf{x}}{N_s}}\;, \nonumber \\
\end{eqnarray}
and the lattice implementation of the left/right handed projection operator $\hat{P}_{L/R}(\mathbf{p})$ takes the following form
\begin{align}\label{eq:proj}
\hat{P}^{\rm lat}_{L/R}(\mathbf{q})= \frac{1}{2}%
\begin{pmatrix}
S_yS_z & \pm  S_x i \tilde{p}_z/p_{\rm curl} & \mp S_x i \tilde{p}_y/p_{\rm curl}\\
\mp S_y i \tilde{p}_z/p_{\rm curl}& S_x S_z & \pm S_y i \tilde{p}_x/p_{\rm curl}\\
\pm S_z i \tilde{p}_y/p_{\rm curl} & \mp  S_z i \tilde{p}_x/p_{\rm curl} & S_x S_y 
\end{pmatrix}\,,
\end{align}
where $p_{\rm curl}^2 = 4 \sum\limits_{i}  \frac{\tan^2 ( \pi q_i/N_i)}{a_s^2}$ denotes the eigenvalues of the lattice curl operator, while $S_i =  \frac{1+e^{-2\pi i q_i/N_s}}{2}$ and $i \tilde{p}_{i}$ defined in Eq.~(\ref{eq:latMomBos}) are the momentum-space representation of the nearest-neighbor lattice averages and backward derivatives in Eqns.~(\ref{eq:Bdefinition},\ref{eq:curlBdefinition}). Based on this decomposition, the occupancy of left/right-handed magnetic field modes is simply given by 
\begin{align}
\label{eq:definitionHelictyperMode}
e^2n^{L/R}_{B}(t,\mathbf{q})= \frac{|ea_s^{1/2}\mathbf{B}^{L/R}(t,\mathbf{q})|^2}{|\tilde{\mathbf{p}}| a_s}\,.
\end{align} 
where $\tilde{\mathbf{p}}$ is the gauge field lattice momentum defined in Eq.~(\ref{eq:latMomBos}).

Since the fermion field operator $\hat{\Psi}(t,\mathbf{x})$ is not gauge invariant a physical definition of the occupation number of left/right handed fermions either requires the implementation of a complicated operator or the specification of an appropriate gauge choice. When extracting fermion spectra, we therefore exploit the residual gauge freedom to fix Coulomb gauge $(\partial_i A^{i}(t,\mathbf{x})=0)$ at the time $t$ of the measurement (see. e.g. \cite{Berges:2013fga}) and calculate the helicity projected occupation numbers of particles according to

\begin{align}\label{eq:occupationnumbers}
&n_{F}^{L/R}(t,E_{\mathbf{q}}>0)  = \\
& \quad \qquad  \frac{1}{N_s^3} \sum\limits_{\mathbf{x}}  \sum\limits_{\mathbf{y}} \langle \,\hat{\Psi}^\dagger(t,\mathbf{x}) \, \phi^{+}_{h,{\mathbf{q}}}(\mathbf{x}) \phi_{h,{\mathbf{q}}}^{+ \dagger}(\mathbf{y}) \, \hat{\Psi}(t,\mathbf{y})\, \rangle\,, \nonumber
\end{align}
and similarly for antiparticles
\begin{align}\label{eq:occupationnumbers}
&n_{F}^{L/R}(t,E_{\mathbf{q}}<0)  = \\
& \quad \qquad  \frac{1}{N_s^3} \sum\limits_{\mathbf{x}}  \sum\limits_{\mathbf{y}} \langle \,\hat{\Psi}^\dagger(t,\mathbf{x}) \, \phi^{-}_{h,{\mathbf{q}}}(\mathbf{x}) \phi_{h,{\mathbf{q}}}^{- \dagger}(\mathbf{y}) \, \hat{\Psi}(t,\mathbf{y})\, \rangle\,, \nonumber
\end{align}
where $\phi^\pm_{h,{\mathbf{q}}} (\mathbf{x})$ are the free helicity wave functions in  \Eq{eq:defmodeswavefunctions}.

\subsection{Extraction of growth rates}
\begin{figure}[t!]
\centering
\includegraphics[width=0.5\textwidth]{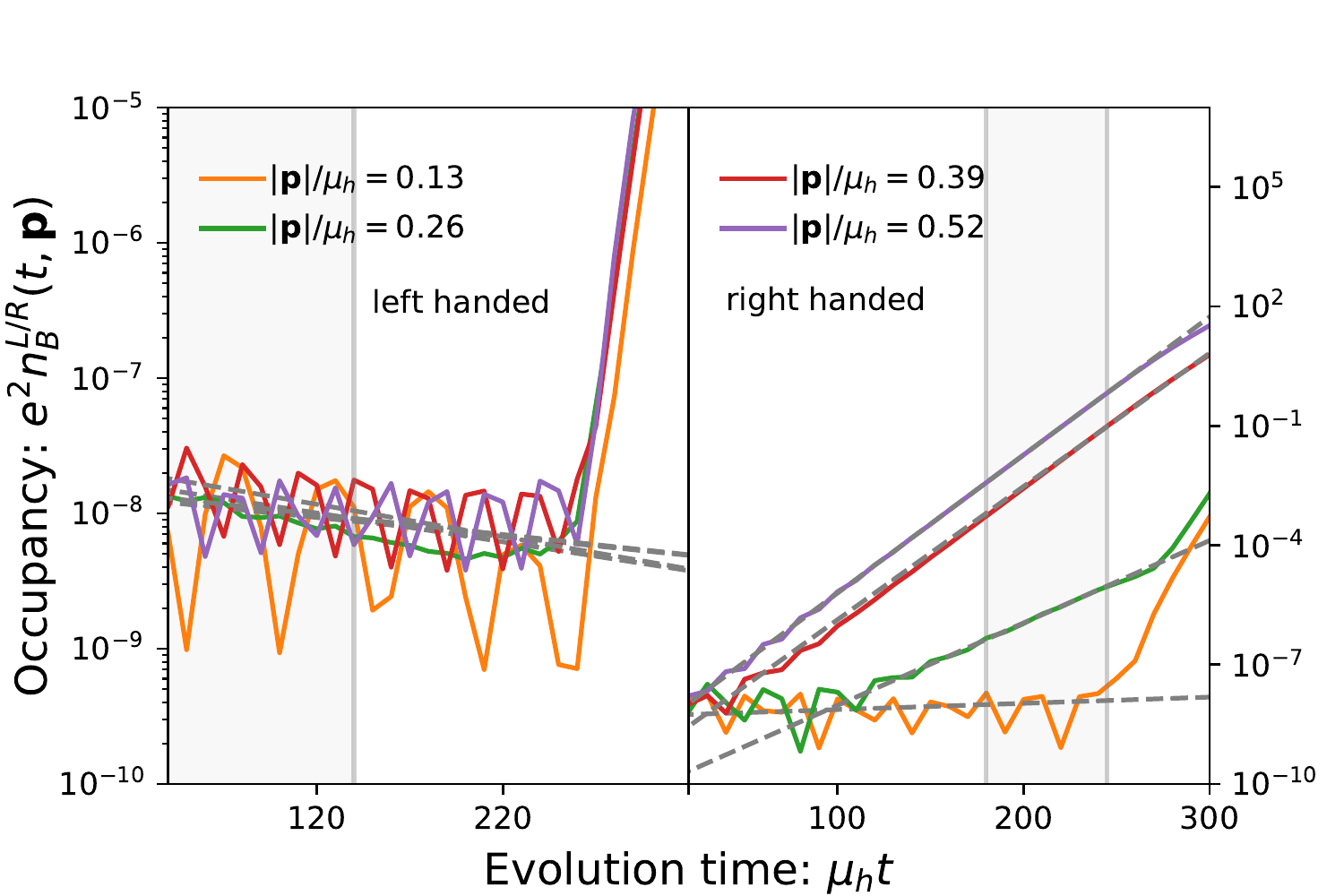}
\caption{Visualization of the extraction of growth and damping rates $\gamma_{L/R}(\mathbf{p})$ from the time evolution of the left/right handed magnetic field modes $e^2n^{L/R}_B(t,\mathbf{p})$. Grey dashed lines represent fits to the functional form Eq.~(\ref{eqn:log_occ}), used to extract damping and growth rates $\gamma_{L/R}$, respectively. Shaded bands indicate the fit region.}
\label{fig:gr_fit}
\end{figure}

When characterizing the exponential growth/damping of left/right handed magnetic fields modes due to the chiral plasma instability, we first extract the evolution of the occupation numbers $e^{2} n^{L/R}_{B}(t,\mathbf{q})$ for each discrete lattice momentum mode $\mathbf{q}$.  When presenting results for the growth of individual momentum modes, as in Fig.~1 of the main text, we average over the different possible orientations by combining modes with the same lattice dispersion $\tilde{\mathbf{p}}$. Subsequently, to extract the growth rates, we perform a linear fit of the form
\begin{eqnarray}
\log\left(e^2n^{L/R}_B(t,\mathbf{p})\right) =\alpha_{L/R}(\mathbf{p}) \mp \gamma_{L/R}(\mathbf{p}) \mu_h t \,,
\label{eqn:log_occ}
\end{eqnarray}
within a fit range which is generally between $\mu_h t \approxeq100-300$, but needs to be determined individually for each momentum mode and lattice configuration by careful inspection to make sure that only the linear instability regime is captured. We illustrate this procedure in Fig.~\ref{fig:gr_fit}, where the extracted growth rates for right-handed modes, $\gamma_R$, and damping rates for left-handed modes, $\gamma_L$, are denoted by the dashed grey lines and the fit range is marked by solid vertical grey lines. Subsequently, we perform an averaging over narrow momentum ($\mathbf{p}$) bins, which determine the horizontal error bars in Fig. 2 of the main text. Since the fit error is marginal the vertical error bars in Fig.~2 of the main text are determined by the statistical uncertainty of growth and damping rates within the respective momentum bin. 

\subsection{Continuum Extrapolation}
Below we provide details on  continuum extrapolation 
of the magnetic helicity ($n_{h}(t)$) and chiral charge  density ($n_{5}(t)$) shown in Fig.~3 of the main text. Based on our simulations, we can immediately infer the time evolution of these quantities
based on their lattice operator definitions Eqs.~(\ref{eq:energydensF}-\ref{eq:energydensB}).  However, when comparing results with different lattice discretizations (i.e. different lattice spacings $a_s$ and volumes $N_s a_s$), 
we find that due to the residual lattice spacing dependence of the instability growth rates, seen in Fig.~2 of the main text, 
the transition to the non-linear turbulent regime, where the dominant chirality transfer from fermions to gauge fields takes place, occurs on slightly different time scales, thereby prohibiting us from directly performing a point-wise continuum extrapolation. 
Hence, in order to compare different data sets at the same stage of the evolution, we first determine the time scale $\tsubsat$ for the transition to the non-linear regime individually for each data set. Based on the criterion that 
at time $t=\tsubsat$ the axial charge density of fermions has decreased by five percent, i.e. $n_{5}(t=\tsubsat)=0.95n_{5}(t=0)$, we obtain the transition times $\tsubsat(a_s)$ given in Tab.~\ref{tab:Offsets}.
Subsequently, we perform point-wise linear and quadratic continuum extrapolations at each time $\mu_{h}(t-\tsubsat(a_s))$, taking into account $5\%$ variations of $\tsubsat(a_s)$ for each data set to account for the uncertainty in this procedure.
Bands of the continuum extrapolation shown in Fig.~3 of the main text correspond to the envelope of linear and quadratic extrapolations (including their errors) and we have checked that the procedure yields stable bands
when varying the offset margins. 
\begin{center}
\begin{table}[]
\begin{tabular}{| c | c | c | c | c | c |}
\hline
$N_s$  & 48  & 48  & 32  & 32  & 32  \\
\hline
$\mu_h a_s$  & 0.666 & 1   & 1   & 1.25 & 1.5 \\
$\mu_h \tsubsat $ & 375 & 305 & 305 & 270  & 240 \\
\hline
\end{tabular}
\caption{\label{tab:Offsets} Extracted time scales $\mu_{h}\tsubsat$ for the transition to the non-linear turbulent regime for different data sets.}
\end{table}
\end{center}

\begin{figure}[t!]
\centering
\includegraphics[width=0.5\textwidth]{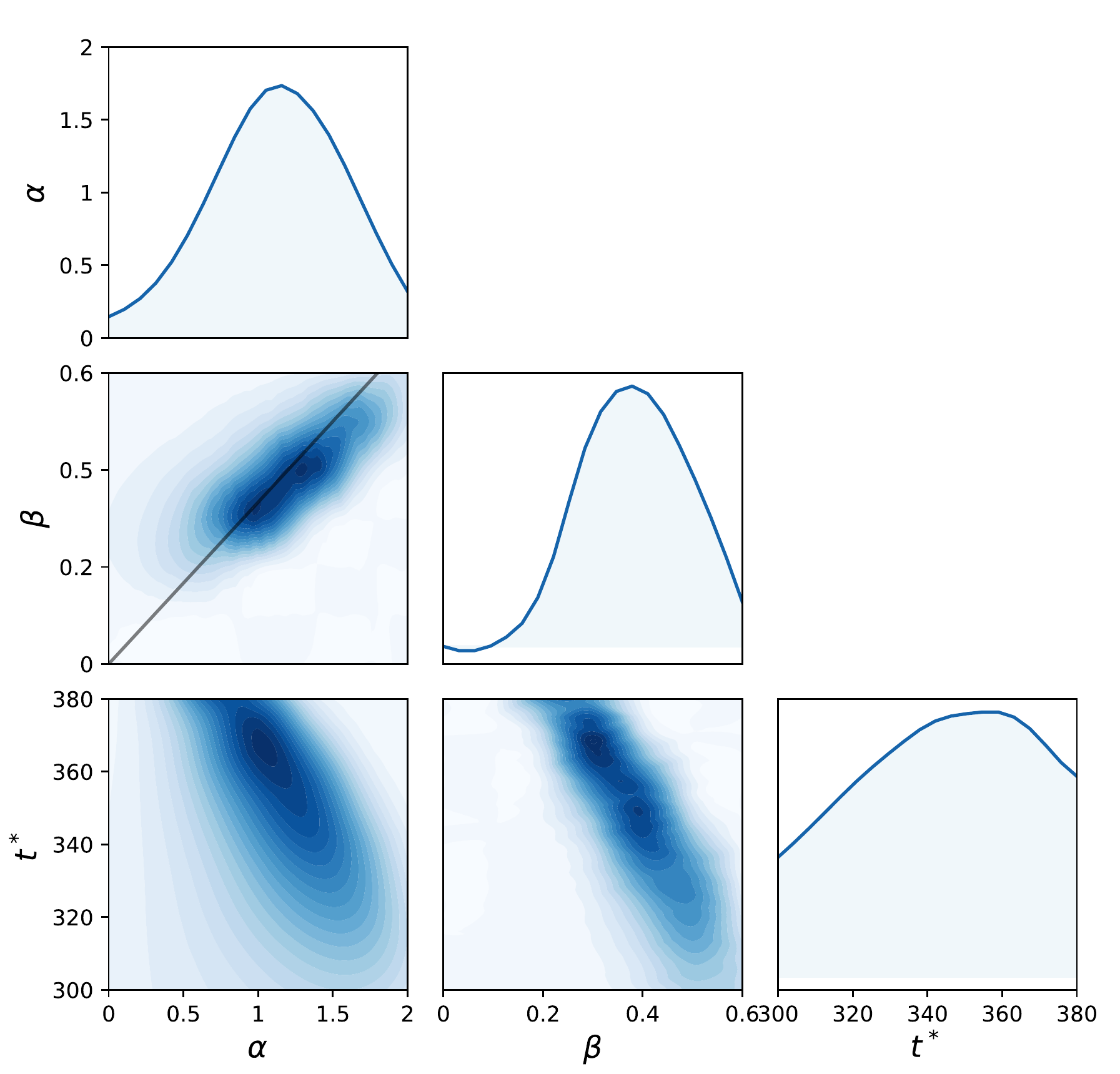}
\caption{Reduced likelihood distributions $W(\alpha)$ (top left), $W(\beta)$ (middle center)  and  $W(t^*)$ (bottom right) are shown along the diagonal; on the off diagonal are the reduced likelihood distribution $W(\alpha,\beta)$ (center left), $W(\alpha,t^*)$ (bottom left)  and $W(\beta, t^*)$ (bottom center). The solid line in the center-left panel shows the scaling relation $\alpha-d\beta=0$ for magnetic helicity conservation.}
\label{fig:likelihoods}
\end{figure}

\subsection{Statistical extraction of scaling exponents}

We perform a statistical analysis of self-similarity following the procedure outlined Appendix D of~\cite{Berges:2013fga} to determine the most likely values of $t^*$, $\alpha$, and $\beta$.  
Because this analysis is independent of lattice discretization, we use continuum notation for the subsequent discussions. Based on the characterization of self-similar behavior of the magnetic helicity spectrum in \Eq{eq:Scaling} of the main text, we first define a reference function 
\begin{eqnarray}
f_{\mathrm{ref}}(t,\hat{p}=\tau_{\mathrm{ref}}^\beta p)= \log \tau_{\mathrm{ref}}^{-\alpha} e^2\Delta n_{B}(\tau_{\mathrm{ref}}^\beta p)\,,
\end{eqnarray}
where $\tau_{\mathrm{ref}} = \mu_h(t_{\mathrm{ref}}-t^*)$. In order to determine the quality of scaling for each combination of $t^*$, $\alpha$, and $\beta$, we compare the reference function to the scaled spectrum at a number of different `test` times, defined analogously as 
\begin{eqnarray}
f_{\mathrm{test}}(t,\hat{p}=\tau_{\mathrm{test}}^\beta p)= \log \tau_{\mathrm{test}}^{-\alpha} e^2\Delta n_{B}(\tau_{\mathrm{test}}^\beta p)\;,
\end{eqnarray}
where $\mu_h t_{\mathrm{test}}=450,475,500,525,550,575,600$ typically. We quantify the deviations from ideal scaling in terms of
\begin{eqnarray}
\chi^2(\alpha, \beta ,t^*)= \frac{1}{N_{\mathrm{test}}} \displaystyle\sum_{\tau_{\mathrm{test}}} \frac{\int \frac{d\hat{p}}{\hat{p}} \left(f_{\mathrm{ref}}(\hat{p})-f_{\mathrm{test}}(\hat{p})\right)^2  }{\int  \frac{d\hat{p}}{\hat{p}}  \left(f_{\mathrm{ref}}(\hat{p})\right)^2}\;,
\end{eqnarray} 
where $N_{\mathrm{test}}$ is the number of `test` times, and the $\tilde{p}$ integration is restricted such that the physical momenta $p$ are in the infrared regime  $0.1<p/\mu_h<0.5$. 
We then quantify the likelihood that a given set of parameters gives the best overall rescaling by the function
\begin{eqnarray}
W(\alpha,\beta,t^*)=\frac{1}{\mathcal{N}} \exp \left[ -\frac{\chi^2(\alpha,\beta,t^*)}{\chi^2_{\mathrm{min}}} \right]\,,
\end{eqnarray}
where $\chi^2_{\mathrm{min}}=1.7\cdot10^{-3}$ denotes the minimal value of $\chi^2(\alpha, \beta ,t^*)$ obtained for $\alpha \approx 0.9$, $\beta \approx 0.3$, $\mu_h t^* \approx 375$ employed in Fig.~\ref{fig:gspectra} of the main text. We visualize the results of the statistical scaling analysis in Fig.~\ref{fig:likelihoods}, where the different panels show the reduced likelihood distributions $W(\alpha,t^{*})$, $W(\beta,t^{*})$ and $W(\alpha,\beta)$ which are obtained by marginalizing over the other variables, e.g. $W(\alpha,t^{*})=\int d\beta~W(\alpha,\beta,t^*)$ and similarly for the other components. Based on the self-similar scaling behavior, the evolution of the net helicity of the system can be evaluated according to
\begin{align}
n_{h}(t) &= e^2\int \frac{d^3 \mathbf{p}}{(2\pi)^3} \Delta n_{B}(\mathbf{p},t) = \tau^\alpha  \int \frac{d^3 \mathbf{p}}{(2\pi)^3}   f_s(\tau^\beta \mathbf{p}) 
\nonumber\\
&= \tau^{\alpha-d \beta} \int \frac{d^3 \hat{\mathbf{p}}}{(2\pi)^3}  f_s(\mathbf{\hat{p}}) \simeq \text{const} \times  \tau^{\alpha-d \beta}\,,
\end{align}
such that for $\alpha-d \beta=0$ the net magnetic helicity is conserved. We find that this scaling relation is approximately satisfied, as indicated by the solid line in the left central panel of Fig.~\ref{fig:likelihoods}.

\end{document}